\documentclass{article}
             
\usepackage{jheppub} % for details on the use of the package, please
\usepackage[T1]{fontenc} % if needed
\usepackage{color}
\usepackage{graphicx}
\usepackage{subfig}
\usepackage{amsmath}
\usepackage{tikz}
\usepackage{caption}
\usepackage{float}

\newcommand{\ba}{\begin{array}}
\newcommand{\ea}{\end{array}}
\newcommand{\bd}{\begin{displaymath}}
\newcommand{\ed}{\end{displaymath}}
\def\be{\begin{equation}}
\def\ee{\end{equation}}
\def\bsube{\begin{subequation}}
\def\esube{\end{subequation}}
\def\bea{\begin{eqnarray}}
\def\eea{\end{eqnarray}}
\def\bal{\begin{align}}
\def\ealign{\end{align}}
\def\eal{\end{align}}
\def\ben{\begin{enumerate}}
\def\een{\end{enumerate}}

\def\beq{\begin{equation}\ba{rcl}}
\def\eeq{\ea\end{equation}}

\def\bsub{\begin{subequations}}
\def\esub{\end{subequations}}

\title{ILC sensitivity for leptophilic scalar dark matter}

\author[a]{Bharti Rawat}

\affiliation[a]{Department of Physics and Astrophysics,\\
                 University of Delhi, Delhi 110007, India.}
\emailAdd{bhartirawat87@gmail.com}
\abstract{We explore the viability of detecting a leptophilic scalar dark matter at the ILC in a simplified model approach. We present the constraints on the couplings of scalar dark matter with the standard model fermions ensuing from the relic density bounds deduced from the 2018 Planck data. We, then, present the reach of the ILC in terms of $\Lambda$, the scale of the effective theory, by performing a detailed analysis for the $Z$ associated DM pair production (i.e., $ e^+ e^- \rightarrow 2 $ jets + $\not \!\! E_T$, $ e^+ e^- \rightarrow \mu^+ \mu^-$ + $\not \!\! E_T$ and $ e^+ e^- \rightarrow e^+ e^-$ + $\not \!\! E_T$). We present the results for various run scenarios as 3-$\sigma$ contours in the $m_{\phi}-\Lambda$ plane. We also analyze the effect of beam polarization on the sensitivity of this search. We find that for the process with two hadronic jets in the final state, ILC can probe $\Lambda$ up to 1.76 TeV for $\sqrt{s}= 1$ TeV that can further enhance to 1.99 TeV, after the inclusion of polarization effects.}

\begin{document} 
\maketitle
\flushbottom
%%%%%%%%%%%%%%%%%%%%%%%%%%%%%%%%%%%%%%%%%%%%%%%%%%%%%%
\section{Introduction}
\label{sec:intro} 
%%%%%%%%%%%%%%%%%%%%%%
%%%%%%%%%%%%%%%%%%%%%%
\par Speculations about the presence of dark matter (DM) in our universe is now well supported by a variety of observational evidence and particulate dark matter has long been a preferred solution to such astrophysical and cosmological discrepancies. The recent results from the Planck 2018 data infer a value of $\Omega_{DM} h^2$ = 0.120 $\pm 0.001$~\cite{Planck:2018vyg}. The corresponding value deduced from the data collected from the 9-year WMAP measurements is $\Omega_{DM} h^2$ = 0.1138 $\pm 0.0045$~\cite{Bennett:2012zja}. These results imply that the total energy density of the universe receives a contribution of approximately 27\% from cold dark matter alone, whereas the contribution of the baryonic matter is only about 4\% of the total energy density. So, the foremost requirement is that the parameters of any proposed theory, along with the early universe conditions and production mechanisms, must give us the correct, observed abundance of dark matter. Cosmological and astrophysical observations also limit the strength of possible electromagnetic interaction between dark matter particles and photons\footnote{While milli-charged DM is still possible~\cite{Liu:2019knx, Caputo:2019tms}, such models tend to be very specific and complicated and we eschew these.}~\cite{McDermott:2010pa,Sigurdson:2004zp,Profumo:2006im}. Similarly, a significant coupling between baryon and dark matter would affect the structure formation. All such considerations renders WIMP one of the most interesting portals to relate the existing extensions of the Standard Model (SM) to the dark matter (DM) problem.
%%%%%%%%%%%%%%%%%%%%%%
\par With a wide variety of DM candidates being possible, the next challenge is to identify the nature of this non-luminous matter. Astrophysical and cosmological observations allow us to infer the total density of dark matter in the universe. However, a precise determination of the dark matter properties and the parameters of the underlying theory requires other detection methods. The first class of experimental signatures are the Direct detection ones where the key idea is to directly detect the local dark matter particles from our Galactic halo scattering off ordinary nuclei. The DAMA/NaI~\cite{Bernabei:2013xsa} Collaboration indeed reported the observation of an annually modulating rate consistent with dark matter scattering. However, their results seem to be inconsistent with the results obtained from subsequent experiments like CoGeNT~\cite{Aalseth:2012if}, XENON100~\cite{Aprile:2016wwo}, PandaX-II~\cite{Yang:2016odq,Tan:2016zwf} and LUX~\cite{Akerib:2016vxi}. These and other experiments have tested and ruled out an impressive range of particle dark matter models by imposing bounds in the plane described by the DM particle mass and its coupling to nucleons. The second possibility is the annihilation of dark matter particles resulting in the production of standard model particles which, in turn, could provide a possible way to detect these particles. This forms the basis of the Indirect detection methods where the focus lies in the observation of cosmic-rays created by WIMP annihilations in galactic halos and the signatures from WIMPs captured in massive bodies such as the sun or stars. While there have been occasional anomalies in the data such as information from gamma-ray satellite missions and the measurement of the cosmic ray antiproton spectrum, the experiments have failed to validate each other’s positive sightings, resulting, once again, in further constraints. 
%%%%%%%%%%%%%%%%%%%%%%
\par The primary focus of the work presented in this paper is the third category of dark matter detection experiments, i.e., particle accelerators. The collision of ordinary standard model particles at tremendous energies could lead to the emergence of undiscovered particles, and this fact forms the basis of collider experiments. We, therefore, look for the possibility of producing DM particles at a future linear collider (the ILC) via processes inverse to those that result in DM annihilation. The discovery of new particles at the present and future colliders and a detailed analysis of their production rates and signals can ultimately answer the inexplicable riddles of particle physics. The existing theoretical models that give us a dark matter candidate include supersymmetry, extra-dimensions, Higgs models, left-right symmetric models, etc. In this publication, we adhere to a simplified model approach and consider different types of effective interactions between scalar DM particles\footnote{The case of leptophilic fermionic DM has been studied extensively, see, for e.g., Refs.~\cite{Dutta:2017ljq,Barman:2021hhg,Kundu:2021cmo}.} and the SM fermions. As we have already argued, the interaction between DM and quarks (nucleons) has stringent constraints from direct detection experiments. On the other hand, the small cross section for DM-electron scattering renders such experiments insensitive to DM-lepton couplings. Seeking to explore this sector, we will, for the sake of convenience, assume that the DM is leptophilic (with vanishingly small tree-level couplings to quarks). This is not a gross over-simplification as such models can be realised easily. This assumption also negates most of the bounds from the Large Hadron Collider (LHC)~\cite{Bhattacherjee:2012ch,Goodman:2010ku,Carpenter:2012rg,Petrov:2013nia,Carpenter:2013xra,Berlin:2014cfa,Lin:2013sca,Bai:2010hh,Birkedal:2004xn,Gershtein:2008bf,Fox:2011fx,Bai:2012xg,Crivellin:2015wva,Petriello:2008pu}.
%%%%%%%%%%%%%%%%%%%%%%
\par This paper comprises five sections. We discuss the effective theory formalism for a leptophilic scalar dark matter in Sec.~\ref{sec:scalardm}. The first step is to construct the effective lagrangian describing the interactions of our DM candidate with SM fermions. We employ the current cosmological bounds on cold dark matter density to calculate and present the limits on the couplings. Here, we consider two possibilities where the dark matter couples to leptons, and all SM fermions, and we present the limits on the couplings for these two scenarios. We begin Sec.~\ref{sec:processes} with a brief discussion of the direct detection and indirect detection and we then proceed to the collider analysis of scalar DM. We exhibit the normalized 1-D differential distributions of various kinematic observables obtained after performing a full simulation of the $Z$-associated DM pair production and the corresponding SM background processes in the context of International Large Detector (ILD). We include the parton showering and hadronization effects and adopt appropriate selection cuts for further analysis. We perform the $\chi^2$ analysis with the differential distribution for the most sensitive observable and present the reach of the ILC in terms of the scale ($\Lambda$) of the effective theory governing the DM-SM fermion interactions in Sec.~\ref{sec:analysisandresults}. We also incorporate the effect of beam polarization. We conclude this paper by summarising our analysis and results in Sec.~\ref{sec:summary}.
%%%%%%%%%%%%%%%%%%%%%%%
%%%%%%%%%%%%%%%%%%%%%%%

\section{Scalar Dark Matter: Relic Density and Constraints on $\Lambda$}
\label{sec:scalardm}
%%%%%%%%%%%%%%%%%%%%%%%%%
As previously mentioned, we focus our attention on a real\footnote{The results for a complex scalar would be exactly analogous. The effective lagrangian describing the scalar and pseudo-scalar interactions of a complex scalar DM will have an additional factor of $1/4$. For a detailed study of MeV scale complex scalar DM, see Ref.~\cite{Choudhury:2019tss}.} scalar DM, denoted as $\phi$ and adopt the effective field theory approach for defining 4-particle effective interactions of the type $\phi$-$\phi$-$f$-$\bar{f}$ between the WIMPs and the SM fermions. In this analysis, we do not consider the interactions involving $\phi$ and the SM gauge or Higgs bosons. We further assume that our WIMP candidate ($\phi$) is the only new particle species at the electroweak scale, and that, any other new particle species is much heavier than $\phi$. Under this assumption, the thermal relic density is entirely determined by the abundance of $\phi$ particles.\footnote{Otherwise, the presence of another DM component having a mass nearly of the order of $\phi$ could significantly alter the relic density of dark matter particles~\cite{PhysRevD.43.3191}.} The stability of the DM is ensured by imposing a $Z_2$ symmetry (under which the DM field is odd, while all the SM fields are even), implying that the interaction terms in the lagrangian have to involve a pair of DM fields. Thus, the effective lagrangian\footnote{One would expect a term like $g_{\phi\,H}\,\phi\,\phi\,H^{\dagger}\,H$. For $m_{\phi} < m_H/2$, the bounds from the invisible higgs decay width i.e., $H\rightarrow \phi\,\phi$, require $g_{\phi\,H} \lesssim 0.004$. Also, for light $\phi$, one has a very efficient higgs-mediated annihilation into a $b\,\bar{b}$ pair. Retaining sufficient relic density requires $g_{\phi\,H} \lesssim 0.001$ for such $m_{\phi}$. On the other hand, for heavier $\phi$, a $\phi\,\phi\rightarrow H\,H$ annihilation can be efficient for $g_{\phi\,H} \gtrsim 0.01$. Since this is well-studied, in our analysis we assume that $g_{\phi\,H}$ is negligibly small.} for the scalar and pseudo-scalar interactions can be written as
%%%%%%%%%
\begin{equation}
{\mathcal L}_{\rm int} \equiv \frac{\text{v}\,\phi^2}{\Lambda^2} \sum_f \, \left[ \bar{f} \left (  g_S^f + {\rm i} \, g_P^f \, \gamma_{5} \right ) f \right]
\label{lag}
\end{equation}
%%%%%%%%%%%
where v is the vacuum expectation value of the Higgs field, $g_{S/P}^f$ is the corresponding strength for the scalar/pseudo-scalar interaction and $\Lambda$ refers to the cut-off scale of the effective
theory.
%%%%%%%%%%%
%%%%%%%%%%%%%%%%%%%%%%%%%%%%%%%%%%%%%%%%%%%%%
\subsection{Relic Density of Scalar DM}
\label{sec:relicdensity}
%%%%%%%%%%%%%
\par Before we proceed to the analysis of collider simulation, we must calculate the current cosmological bounds on our DM candidate. For this purpose, we present the relic density calculations which involve a simple idea that, in the early universe, interactions among particles were very frequent compared to the later epochs. As a consequence of multiple and frequent interactions, the constituents of the universe were in equilibrium. The equilibrium abundance is maintained by annihilation of the particle with its antiparticle into lighter SM particles $l$ $\left ( \phi \phi \rightarrow l \bar{l} \right )$ and vice versa $\left (l \bar{l}\rightarrow \phi \phi \right )$. As the universe cools to a temperature less than the mass of the particle, the inverse process is kinematically suppressed and the equilibrium abundance drops exponentially as $e^{-m_{\phi}/T}$. And, when the rate for the annihilation reaction falls below the expansion rate $H$, the interactions which maintain thermal equilibrium cease. This is known as ``freeze out''. This entire phenomena is explained by the Boltzmann equation given as
%%%%%%%%%%%%
\begin{eqnarray}
a^{-3}\,\frac{d}{dt}\,\left(n_{\phi}\,a^{3}\right)= \left \langle \sigma\,v \right \rangle \left[n_{\phi}^{(0)2}-n_{\phi}^{2} \right],
\end{eqnarray}
%%%%%%%%%%%
where, $a$ is the scale factor, $n_{\phi}^{(0)}$ is the equilibrium number density of $\phi$ and $\left \langle \sigma\,v \right \rangle$ is the thermal average of the annihilation cross section times the relative velocity $v$ (of a pair of $\phi$'s).
\par Solving the Boltzmann equation gives us the following result for the fraction of energy density contributed by the dark matter 
%%%%%%%%%%
\begin{eqnarray}
\Omega_{\phi} = \sqrt{\frac{4\,\pi ^3\,G\,g_{\star}(m_{\phi})}{45}}\frac{x_f\,T_0^3}{30\,\left \langle \sigma\,v \right. \rangle\,\rho_{cr}}.\label{omega}
\end{eqnarray}
%%%%%%%%%%
Here, $x_f \equiv m_{\phi}/{T_f}$ with $T_f$ being the freeze-out temperature, $g_{\star}(m_{\phi})$ gives us the effective degrees of freedom at freeze-out, $T_0$ is the present CMB temperature with a value $2.72548 \pm 0.00057 \,\rm K$~\cite{Fixsen:2009ug} and 
$\rho_{cr} = 1.05375 \times 10^{-5} \, h^2 \left( GeV/c^2 \right) cm^{-3}$ is the critical density of the universe.
We can re-write equation~\eqref{omega} as 
%%%%%%%%
\begin{eqnarray}
\Omega_{\phi}\,h^2 = 1.04 \times 10^{9}\, {\rm GeV^{-1}} \frac{x_f}{m_{Pl}\,\sqrt{g_{\star}(T_f)}\,\left \langle \sigma\,v\right \rangle}.
\label{omega1}
\end{eqnarray}
%%%%%%%%
Here $m_{Pl}$ is the Planck mass. The relation above tells us that $\Omega_{\phi}$ depends mainly on the annihilation cross section of the dark matter particles. Therefore, to determine the relic density of our scalar DM, we need to calculate the cross section for DM pair-annihilation to a pair of SM fermions ($f$). For the effective interactions mentioned in~\eqref{lag}, the cross sections are given by
%%%%%%%%%%%
\begin{subequations}
\begin{eqnarray}
\sigma^{\rm ann}_{S} & = &
\frac{\text{v}^2}{2 \, \pi \, \Lambda^4} \, \sum_f \, \left(g^f_{S}\right)^2 {C}_f \,
\sqrt{\frac{s-4 \,m_f^2}{s-4 \,m_{\phi}^2}} \,
\left( \frac{s-4\,m_f^2}{s} \right),\label{sigma_ann_S}\\
\sigma^{\rm ann}_{P} & = & 
\frac{\text{v}^2}{2 \, \pi \, \Lambda^4} \, \sum_f \, \left(g^f_{P}\right)^2 {C}_f \,
\sqrt{\frac{\rm s-4 \,m_f^2}{s-4 \,m_{\phi}^2}} \, ,
\label{sigma_ann_P}
\end{eqnarray}
\end{subequations}
%%%%%%%%%%%
where ${C}_f$ denotes the color factor, equal to 1 for leptons and 3 for quarks.
%%%%%%%%%%%%%%%%%%%%%%%%%%%%%%%%%
\subsection{Constraints on $\Lambda$ from Cosmological Observations}
\label{sec:sv}
%%%%%%%%%%%%%%%%%
\par The information of the rate at which a DM particle will get annihilated with another is embedded in the thermally-averaged value of $\langle \sigma^{\rm ann}_{S/P} \,v \rangle$. Using equations~\eqref{sigma_ann_S} and~\eqref{sigma_ann_P}, we now proceed to calculate the quantity $\sigma^{\rm ann}_{S/P} \,v$\footnote{For convenience, we perform this calculation in the rest frame of one of the initial particles, with $v$ being the velocity of the second particle in this frame.}. Since our DM $\phi$ is a cold relic\footnote{To carry out this expansion, we follow the approach mentioned in Ref.~\cite{GONDOLO1991145}.}, we can safely assume that it was non-relativistic at the epoch of decoupling. Under this assumption, we can parameterize the annihilation cross section times the relative velocity as,
%%%%%%%%%%%
\begin{eqnarray} 
\sigma_{ann}\,v &\simeq& a + b\,v^2 +\mathcal{O}(v^4). 
\end{eqnarray}
%%%%%%%%%%%
The expansion $\sigma \, v$ can be carried out by expressing the mandelstam variable, $s$, in terms of the velocity as, 
%%%%%%%%%%%
\begin{eqnarray}
s &\simeq& 4\,m_{\phi}^2+m_{\phi}^2\,v^2+ \frac{3}{4} \, m_{\phi}^2\,v^4+\mathcal{O}(v^6).
\end{eqnarray}
%%%%%%%%%%%%%
Substituting this expansion in equations~\eqref{sigma_ann_S} and~\eqref{sigma_ann_P}, we get the following expressions (to $\mathcal{O}(v^2)$),
%%%%%%%%%%%%%
\begin{subequations}
	\begin{eqnarray}
	\sigma^{\rm ann}_{S} \, v &\simeq& 
	\frac{\text{v}^2}{\pi\, \Lambda^4} \,
	\sum_f \left(g^f_{S}\right)^2 \, {C}_f \, \sqrt{1-\frac{m_f^2}{m_{\phi}^2}}
	\left[ \left( 1- \frac{m_f^2}{m_{\phi}^2} \right) + \frac{1}{8} \left( -2 + 5 \frac{m_f^2}{m_{\phi}^2} \right ) v^2 \right], \label{sigmav_S}\\
	\sigma^{\rm ann}_{P} \, v &\simeq& 
	\frac{\text{v}^2}{\pi \, \Lambda^4} \,
	\sum_f \left(g^f_P\right)^2 \, C_f \, \sqrt{1-\frac{m_f^2}{m_{\phi}^2}}\,
	\left [ 1 + \frac{-2 + 3 \frac{m_f^2}{m_{\phi}^2}}{8 \, \left ( 1-\frac{m_f^2}{m_{\phi}^2} \right )} v^2 \right ]. \,
    \label{sigmav_P}
	\end{eqnarray}
\end{subequations}
%%%%%%%%%%%%%%%
The terms which are independent of $v$ emerge from $s$-wave scattering, while the ${\mathcal O}(v^2)$ terms receive contributions from both $s$-wave and $p$-wave annihilation processes.
%%%%%%%%%%%%%%%%%%%%
%%%%%%%%%%%%%%%%%%% 
\par Assuming (for the time being) $\phi\,\phi \rightarrow e^+ \, e^-$ to be the dominant channel into which the $\phi$'s annihilate, we can estimate the quantity $\langle\sigma^{\rm ann}_{S/P} \,v\rangle$. At present, the observational bounds on the cold dark matter (CDM) density from the 2018 Planck data is $\Omega_{CDM} h^2$ = 0.120 $\pm 0.001$~\cite{Planck:2018vyg}. As is evident from Eq.~\eqref{omega1}, this limit translates to a lower bound on the quantity $\left \langle \sigma\,v \right \rangle$. This, in turn, would imply an upper bound on the cut-off scale ($\Lambda$) of our effective theory. Therefore, Eq.~\eqref{omega1} together with Eq.~\eqref{sigmav_S} and Eq.~\eqref{sigmav_P} gives us the bounds on $\Lambda$ for the range of scalar DM mass relevant for our analysis. 
%%%%%%%%%%%%%%%%%%%%%%
\par For simplicity, we consider the scalar and pseudo-scalar interactions separately and for each interaction(S/P), we present the bounds corresponding to the following scenarios
%%%%%%%%%%%
\begin{itemize}
	\item[$\bullet$] In the first case, we consider a leptophilic scalar dark matter where the DM couples to leptons alone. We further assume that the DM-lepton couplings are universal in nature i.e. $g_{S/P}^l = 1$, for all leptons. 
	
	\item[$\bullet$] In the second case, we consider the possibility where dark Matter couples to all standard model fermions (SMF). In this scenario also, the DM couplings to all SMF are universal i.e. $g_{S/P}^f = 1$, for all SMF\footnote{We present the bounds on $\Lambda$ corresponding to this scenario only for reference. This is not relevant for the linear collider analysis discussed in the subsequent sections.}
\end{itemize}
%%%%%%%%%%%%%%%%	
\par
\begin{figure*}[htb]
	\centering
	\begin{tabular}{cc}
		\includegraphics[width=7cm,height=9cm,angle=-90]{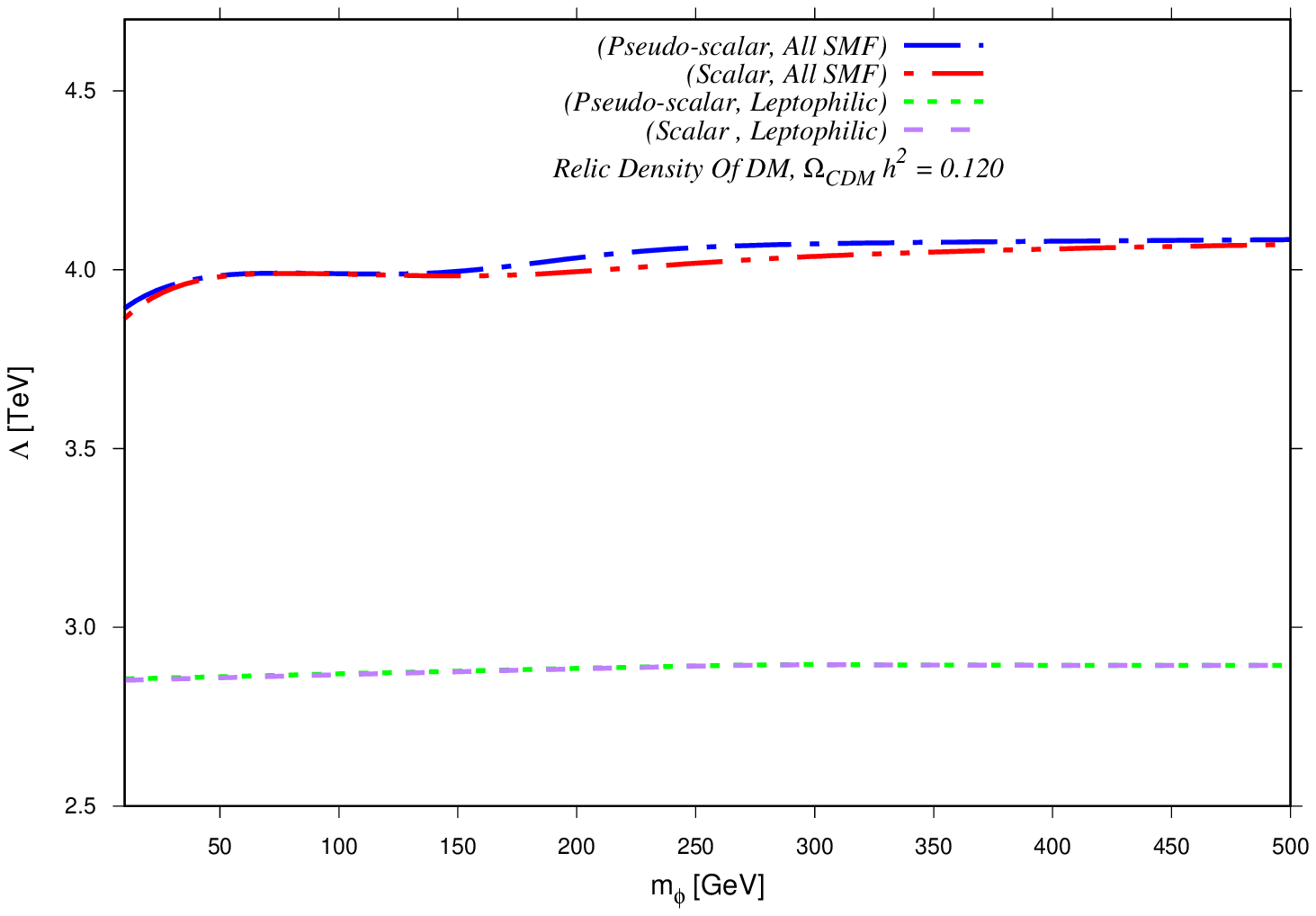} &
	\end{tabular}
	\caption{\small \em{Contours in the $\Lambda$--$m_{\phi}$ plane based on the cold dark matter density $\Omega_{CDM} h^2$ = 0.120 $\pm 0.001$ obtained from the 2018 Planck data~\cite{Planck:2018vyg}. These curves are obtained using $micrOmegas \mbox{-} v \mbox{-}5.2.1$~\cite{Belanger:2018ccd}, the regions below the curves are allowed by the relic density bounds.}}
	\label{fig:relicdensity} 
\end{figure*}

 Next, we calculate the relic density using micrOmegas~\cite{Belanger:2018ccd}. For this purpose we incorporate the effective interactions described by Eq.~\eqref{lag} via Feynrules2.0~\cite{Alloul:2013bka}. We present the relic density plots showing the upper bounds on $\Lambda$ in Fig.~\ref{fig:relicdensity}. These curves can be interpreted as follows
%%%%%%%%%%%%%%%%%%%%%%
%%%%%%%%%%%%%%%%%%%%%%
\begin{itemize}
	\item[$\bullet$] The enormous difference in the $\Lambda$ values for the two scenarios results from the fact that the inclusion of all standard model fermions will lead to a higher self-annihilation cross section. In this case, for $\Omega_{\phi} \, h^2$ values to be consistent with the current cosmological observations, higher $\Lambda$ values would be required.

	\item[$\bullet$] For the two scenarios, both scalar and pseudo-scalar interactions give nearly identical $\Lambda$ values. The difference becomes apparent only for the scenario involving quarks. This behaviour can be understood from Eqs.~\eqref{sigmav_S} and~\eqref{sigmav_P}, which clearly show that the quantity $\sigma^{\rm ann}_{S/P} \,v \propto 1/\Lambda^2$ and the dependence on $m_{\phi}$ arises only through other terms having a factor of $m_f^2/m_{\phi}^2$. However, for the DM mass we consider, this would only reflect in the case of annihilation into a top pair. Therefore, the curve for scalar interactions show a slight deviation (drop) around $m_{\phi} \simeq m_{top}$.
\end{itemize}
%%%%%%%%%%%%%%%%%%%%%%%%%%%
%%%%%%%%%%%%%%%%%%%%%%%%%%%
%\par \textcolor{red}{While scalar dark matter scenarios have been previously explored in the context of LHC~\cite{Bhattacherjee:2012ch,Goodman:2010ku,Carpenter:2013xra,Fox:2011fx}, we are more interested in the case of the DM coupling primarily to the SM leptons and assume, $g^q_{i} = 0$, throughout this analysis. We present the detailed analysis in the next section.}
%%%%%%%%%%%%%%%%%%%%%%%%%%%%%%
%%%%%%%%%%%%%%%%%%%%%%%%%%%

\section{DM Detection Methods}
\label{sec:processes} 
For a leptophilic dark matter, as the DM--quark couplings are not allowed, the tree-level scattering between dark matter and the nucleus cannot occur. Therefore, DM-nucleus interaction is only possible via higher-order loop diagrams containing a virtual lepton pair that couples electromagnetically to quarks. For both scalar and pseudo-scalar couplings, the one-loop contribution vanishes. Then for scalar couplings, the lowest order contribution to DM--quark interaction comes from the two-loop diagrams\footnote{This has been discussed rigorously in Ref.~\cite{Kopp:2009et}. We refer to their results to draw some significant conclusions.}, i.e., to obtain non-vanishing results from the traces inside the loop integrals, the scalar currents must have at least two photons. This contribution suffers a suppression by a factor of $\sim\,(\alpha_{em}/4\,\pi)^2$ and is therefore negligible. For pseudo-scalar couplings, the diagrams vanish to all loop orders.
 
\begin{figure}[h!]
	\centering
	\includegraphics[width=7cm,height=11cm,angle=-90]{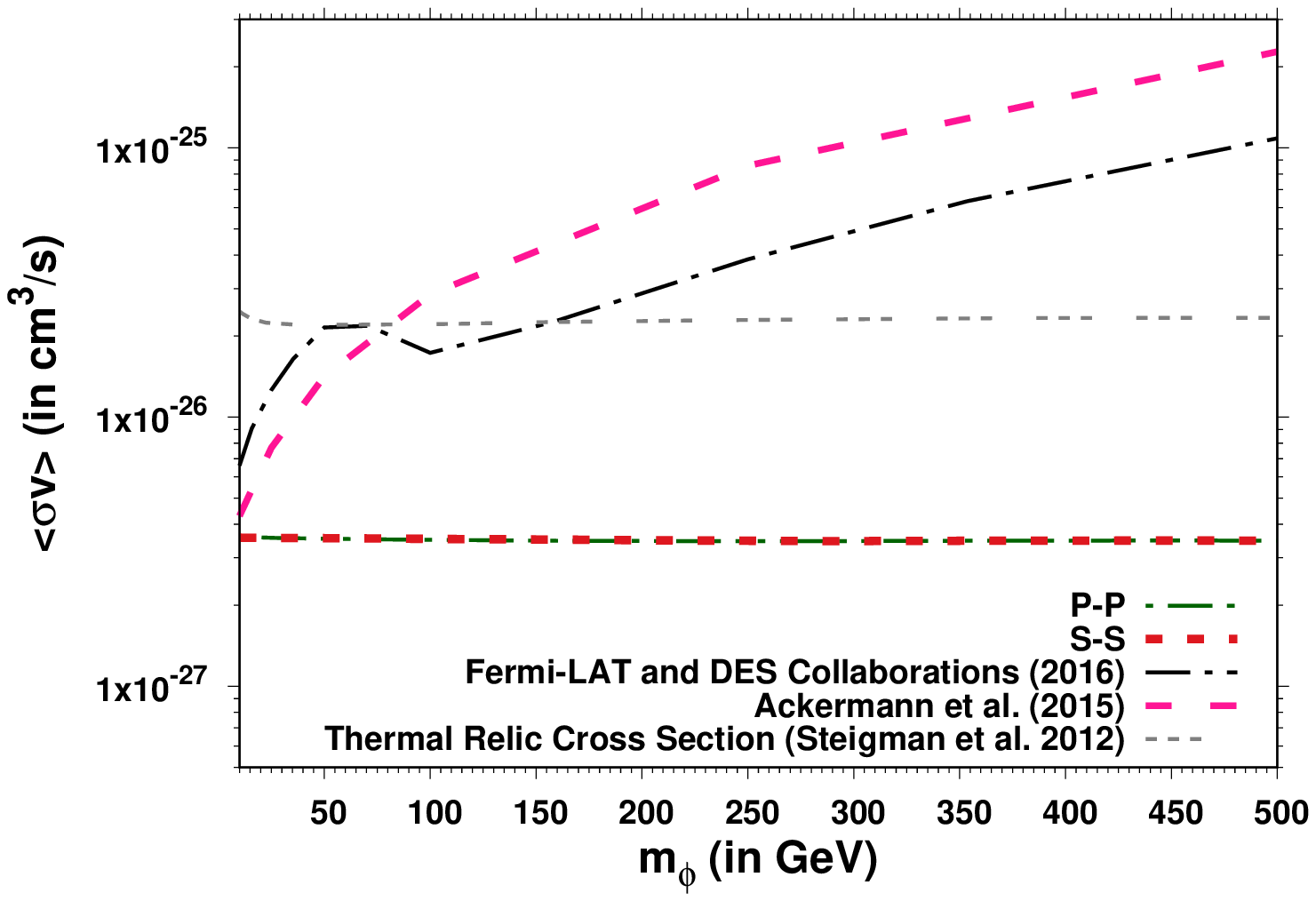} 
	\caption{\small \em{Comparison of the thermally averaged DM annihilation cross sections using the relic density allowed lower bound on scalar and pseudo-scalar couplings of leptophilic DM with the DM annihilation cross section derived from a combined analysis of the nominal target sample for the $\tau^+\tau^-$ channels~\cite{Fermi-LAT:2016uux}.}}
	\label{fig:indirectdetection} 
\end{figure}

Now, to discuss the bounds from the indirect detection experiments, we use Eqs.~\eqref{sigmav_S} \&~\eqref{sigmav_P}, where, v is the local velocity\footnote{This would essentially depend on the profile of the DM distribution assumed for analysis.} of the DM particle. In~\cite{Fermi-LAT:2016uux}, the expected sensitivity is given as a limit on the thermally averaged DM annihilation cross section for the $\tau^+\,\tau^-$ channel. For our DM candidate with mass $m_{\phi}$, we calculate the thermally averaged annihilation cross section using the relic-density allowed lower bound on the scalar and pseudo-scalar couplings . We exhibit our results in Fig.~\ref{fig:indirectdetection} along with a comparison with the bounds from the current indirect detection searches~\cite{Fermi-LAT:2016uux}.

\subsection{Collider Production: Interesting Signatures at ILC}

%%%%%%%%%%%%%%%%%%%%%%%%%%%%%%%%%
%%%%%%%%%%%%%%%%%%%%%%%%%%%%%%%%%
%%%%%%%%%%%%%
\par The stability of any DM candidate over cosmological time scales~\cite{DeLopeAmigo:2009dc,Peter:2010sz,Peter:2010xe} directs us to consider only such collider processes which could lead to the pair production of DM particles. Unfortunately, the production of DM particles alone does not leave any visible signature as the $\phi$'s escape the detector undetected. Having an observable signature requires the $\phi$'s to be produced along with some visible particle(s). While the most general search channel for DM production at any linear collider is the final state with an additional photon radiated from one of the incoming leptons, i. e. $e^+ \, e^- \rightarrow \phi \, \phi \, \gamma$, in this paper, we explore the complementary search channel, i. e. $Z$ associated DM pair-production,
%%%%%%%%%
\begin{eqnarray}
e^+ e^- \to  \phi\,\phi+ Z\, \left(Z\to jj, \mu^+ \mu^- , e^+ e^- \right).
\label{eqn:sig}
\end{eqnarray}
%%%%%%%%%
We include the $b$-jet in our definition of $j$ along with $u,d,s,c,b$ and consider it equivalent to any other light quark jet\footnote{For the present work, we refrain from imposing any b-tagging requirements in our analysis.}. Note that while Eq.~\eqref{eqn:sig} suggests that we consider only an on-shell $Z$, in actuality, we include contributions from an off-shell $Z$ as well. The irreducible SM backgrounds corresponding to the aforementioned channels are
%%%%%%%%%
\begin{eqnarray}
e^+ e^- &\rightarrow& 2 \, {\rm jets}/\mu^+ \mu^-/e^+ e^- + \sum_{i} \nu_i\bar{\nu_i},
\label{eqn:bkgd}
\end{eqnarray}
%%%%%%%%%
respectively. Depending on the center of mass energy, polarization of initial leptons and the cuts, the relative strengths of the three search channels can be determined. We note that the neutrino background associated with dijet search channel in~\eqref{eqn:bkgd} emanates majorly from $e^+ \, e^- \rightarrow Z + \gamma^*/Z^*$. The virtual photon (or, $Z$ boson) eventually splits into 2 jets and the on-shell $Z$ goes to $\nu_i\,\bar{\nu_i}$. Sub-dominant contributions arise from $W^+\,W^-$ fusion diagram, where off-shell $W$'s from the electron legs fuse to a quark pair through either a ``s-channel" $\gamma/Z$ or a ``t-channel" quark; also present are contributions wherein a lowest order $e^+\,e^-\rightarrow \nu_e\,\bar{\nu_e}$ is accompanied by a $\gamma^{\star}/Z^{\star}$ radiated off the leptons and going to a quark-pair. We include all relevant contributions to such final state. SM background processes with leptons in the final state~\eqref{eqn:bkgd} will have similar Feynman diagrams.
%%%%%%%%%%%%%%%%%%
\par We employ MadGraph5~\cite{Alwall:2014hca} to generate the Monte Carlo event samples corresponding to the signal and SM background processes. To carry out the parton shower, and hadronization we use PYTHIA8~\cite{Sjostrand:2014zea}. Furthermore, the simulation of the detector response is performed using DELPHES3~\cite{deFavereau:2013fsa}. We use the configuration card corresponding to the International Large Detector (ILD)~\cite{Behnke:2013lya} incorporated within the Delphes framework. The energy smearing parameters of the electromagnetic and the hadronic calorimeter are set according to the technical design report for ILC detectors~\cite{Behnke:2013lya}, as follows
%%%%%%%%%%%%%%%%%
\begin{subequations}
	\begin{eqnarray}
	\frac{\Delta E}{E} &=& \frac{15 \% }{\sqrt{E/ \rm GeV}} \oplus 1\% \label{eqn:pgsparam1}\\ 
	\frac{\Delta E}{E} &=& \frac{50 \%}{\sqrt{E/ \rm GeV}}\label{eqn:pgsparam2}
	\end{eqnarray} 
\end{subequations}
%%%%%%%%%%%%%%%%%
The jet reconstruction using the anti-$k_T$ algorithm~\cite{Cacciari:2008gp} with a jet cone radius $R = 0.5$ is performed using the FastJet~\cite{Cacciari:2011ma,Cacciari:2005hq} package integrated within the Delphes framework. While performing the simulations, we refer to the technical design report for ILC~\cite{Behnke:2013xla} to set the accelerator parameters and we tabulate them in Table~\ref{table:accelparam} for convenience.
%%%%%%%%%%%%%%%%%%%
\begin{table*}[!h]\footnotesize
	\centering
	\begin{tabular*}{\textwidth}{c|@{\extracolsep{\fill}} cccccc|}\hline\hline
		&\textit{ILC-500}&\textit{ILC-500P}&\textit{ILC-1000}&\textit{ILC-1000P}\\
		$\sqrt{s} \left( \textit{in GeV}\right )$& $500$ & $500$ & $1000$ & $1000$ \\
		$L_{int} \left( \textit{in $fb^{-1}$}\right )$& $500$ & $250$ & $1000$ & $500$ \\
		$P_{-}$        & 0 & 80\% & 0 & 80\%\\
		$P_{+}$        & 0 & 30\% & 0 & 30\%\\\hline\hline
	\end{tabular*}
	\caption{\small \em{Accelerator parameters for each of the run scenarios considered in this paper.
			$P_{-}$ and $P_{+}$ represent the electron and positron polarizations respectively. The table has been taken from~\cite{Dutta:2017ljq}~\cite{Behnke:2013xla}.}}
	\label{table:accelparam}
\end{table*}
%%%%%%%%%%%
%%%%%%%%%%%%%%%%%%%
\subsection{Basic Cuts}
\begin{figure*}[tbh]
	\centering
	\includegraphics[width=7.2cm,height=6cm]{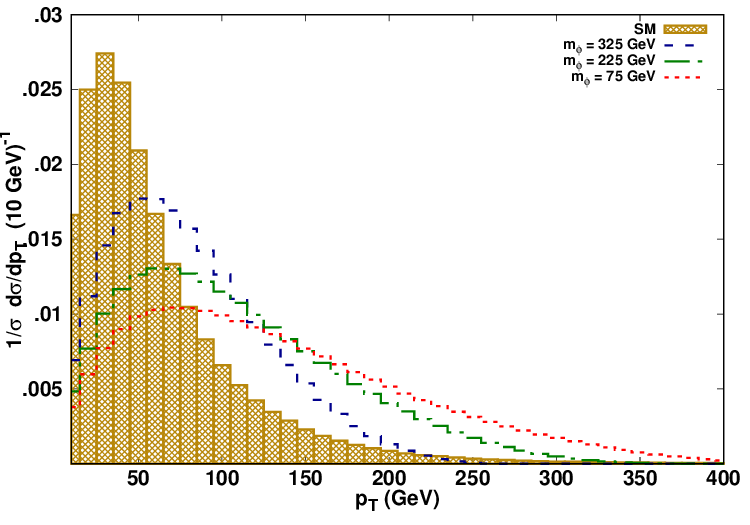}    
	\includegraphics[width=7.2cm,height=6cm]{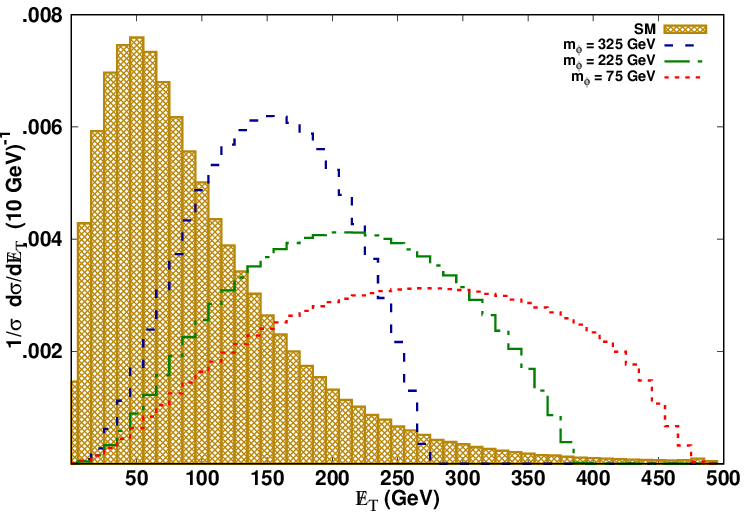}
	\includegraphics[width=7.2cm,height=6cm]{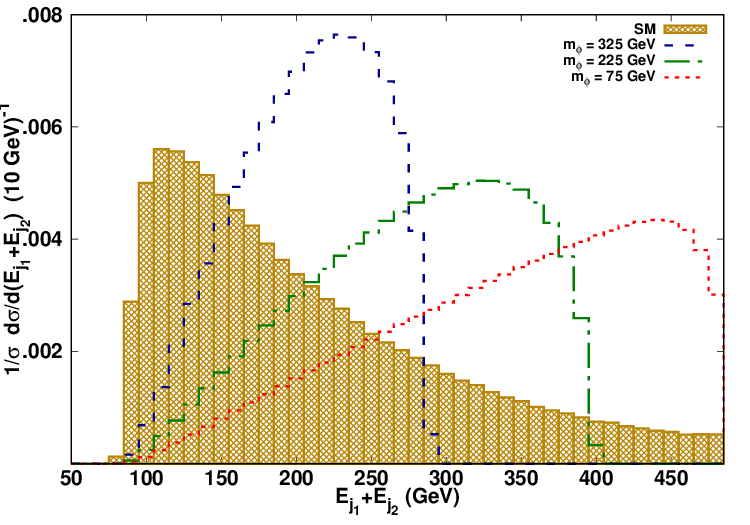}        
	\includegraphics[width=7.2cm,height=6cm]{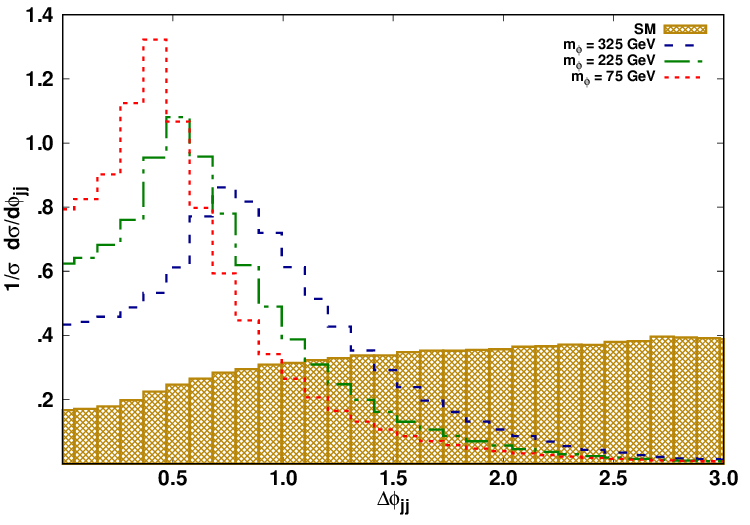}
	\caption{\small \em{Normalized 1-D differential distribution
			of the cross section {\it w.r.t.} kinematic observables
			${p_T}_{j}$, $\not \!\! E_T$, sum of energy of jet pairs and
			$\Delta\phi_{jj}$ for the background process (shaded
			histograms) and the $Z\to j\, j$ associated DM pair production
			at the three representative values of $m_{\phi}$ = 75, 225 and
			325 GeV with $\sqrt{s}$ = 1 TeV, $\Lambda$ =1 TeV and
			$g_{S}^l$ =1. The bin width is 0.1 for the $\Delta\phi_{ii}$ distribution while it
			is 10 GeV for ${P_T}_{j}$, $\not \!\! E_T$ and $E_{j_1}+E_{j_2}$.}}
	\label{fig:distrUS_1TeV}
\end{figure*}

To incorporate the limitations of a realistic collider in our simulations, we begin by implementing certain basic cuts based on the detector characteristics and the theoretical considerations (such as jet definitions). The kinematic variables relevant to our analysis are the transverse momenta ($p_{T j/\ell}$), rapidities ($\eta_{j/\ell}$) and separation ($\Delta R_{jj/\ell\ell}$) of the visible entities (jets and $\mu^\pm$, $e^\pm$) and we demand
%%%%%%%%%%%% 
\begin{itemize}
	\item  $p_{T_{i}} \geq$10 GeV where $i=\ell,j$ ,  
	
	\item  $\left\vert\eta_\ell\right\vert\leq$ 2.5 and $\left\vert\eta_j\right\vert \leq$ 5,
	\item  $\Delta R_{ii} \geq 0.4$ where $i=\ell,j$ \ ,
	\item  $\left\vert M_{ii}-m_Z\right\vert\le 5\,\Gamma_Z$ where $i=\ell,j$; for the hadronic channel, this invariant mass actually refers to that for all the jets together,
\end{itemize}
%%%%%%%%%%%%
\begin{table*}[tbh]\footnotesize
	\begin{center}
		\begin{tabular*}{\textwidth}{c|@{\extracolsep{\fill}} c||c|c|c||c|c|c}\hline\hline
			&&\multicolumn{3}{c||}{\bf \underline{Unpolarized}}&\multicolumn{3}{c}{\bf \underline{Polarized}}\\
			$\left(P_{e^-},\, P_{e^+}\right)$					&	&(0,\,0)	&(0,\,0)	&	&(.8,\,.3)	&(.8,\,.3)	&\\\hline\hline

			&	$m_{\phi}$		&	$\sigma_{sig}$		&	$\sigma_{bkgd}$		&Signal&	$\sigma_{sig}$		&	$\sigma_{bkgd}$		&Signal\\
			& $\text{( in GeV )}$ & $\text{( in pb )}$		& $\text{( in pb )}$		&significance	& $\text{( in pb )}$		& $\text{( in pb )}$		&significance\\\hline\hline      
			&	75	&$1.10 \times 10^{-2}$&	&1.99&$1.37 \times 10^{-2}$&	&7.98\\
$e^+ e^- \rightarrow 2 \, {\rm jets}  + \phi \phi$&	225	&$4.67 \times 10^{-3}$&$0.536$&0.86&$5.79 \times 10^{-3}$&$0.153$&3.54\\
			&	325	&$1.32 \times 10^{-3}$&	&0.24&$1.64 \times 10^{-3}$&	&1.03\\\hline
			&	75	&$5.45 \times 10^{-4}$&	&1.35&$6.76 \times 10^{-4}$&	&4.25\\
$e^+ e^- \rightarrow \mu^+ \mu^- + \phi \phi$&	225	&$2.32 \times 10^{-4}$&$0.035$&0.58&$2.32 \times 10^{-4}$&$0.011$&1.86\\
			&	325	&$6.68 \times 10^{-5}$&	&0.20&$8.28 \times 10^{-5}$&	&0.54\\\hline
			&	75	&$5.45 \times 10^{-4}$&	&1.35&$6.75 \times 10^{-4}$&	&4.25\\
$e^+ e^- \rightarrow e^+ e^- + \phi \phi$&	225	&$2.88 \times 10^{-4}$&$0.035$&0.58&$2.88 \times 10^{-5}$&$0.011$&1.86\\
			&	325	&$6.66 \times 10^{-5}$&	&0.2&$8.25 \times 10^{-5}$&	&0.54\\\hline\hline
			
		\end{tabular*}
	\end{center}
	\caption{\small \em{Significance $S/\sqrt{(S+B) + \delta^2_{sys} (S+B)^2}$ $\left(\rm with\,\,\delta_{sys} = 1 \% \right)$ based on cross sections obtained after the implementation of basic cuts for $\Lambda = 1$ TeV at $\sqrt{s} = 1$ TeV and an integrated luminosity ${\cal L} = 1000 \, \text{fb}^{-1}$ for both unpolarized and polarized initial beams.}}
	\label{table:cs}
\end{table*}
%%%%%%%%%%%%%%%%%%%%%%%%%%%%%
%%%%%%%%%%%%%%%%%%%%%%%%%%%%%
In Table~\ref{table:cs}, we present the cross-section values obtained after the imposition of these basic cuts. We also present a crude estimate of the efficiency of the signal processes mentioned in Eq.~\eqref{eqn:sig} for both polarized and unpolarized initial beams. By invoking the invariant mass cut $\left\vert M_{ii}\right\vert$, we essentially reduce the background by restricting the $Z + Z^*$ final state so that the on-shell $Z$ decays to jets (or, leptons) and the virtual $Z^*$ goes to neutrinos. The SM background with lepton final states will be identically suppressed. This effect would manifest itself in the total cross section rates for the three processes. The normalized distributions, however, are similar. Therefore, for further analysis, we have displayed the one dimensional normalized differential cross section only for process with jets in the final state in Fig.~\ref{fig:distrUS_1TeV}.
%%%%%%%%%%%%%
%%%%%%%%%%%%%%%
%%%%%%%%%%%%%
\subsection{Analysis with one-dimensional Histograms}
\label{sec:analysis}
%%%%%%%%%%%%%%%%
\par To enhance the sensitivity of our analysis, we study the normalized 1-D differential distribution of the cross section {\it w.r.t.} all the sensitive kinematic observables {\it e.g.} $p_{T_{j/\,l}}$, $\not \!\!\! E_T$, $\Delta\phi_{ii}$ and sum of energies of a jet or lepton pair $\left(E_{i_1}+E_{i_2}\right)$. The idea is to select the additional cuts, dubbed "Selection Cuts", in such a way that a substantial contribution from the SM background is eliminated. Therefore, for each observable, the values for which the signal contribution has an excess over SM background are examined for choosing the appropriate selection cuts. Keeping $\sqrt{s}$ = 1 TeV and $\Lambda$ = 1 TeV, we provide a simple yet efficient analysis by choosing our selection cuts such that they apply to all the three channels\footnote{The signal event topologies would be very similar for $\mu^+\,\mu^-$ and dijet final states. While a difference would appear for the $e^+\,e^-$ final state (thanks to the t-channel diagram), the strong cuts on $\left\vert\eta_\ell\right\vert$ and $\left\vert M_{ii}\right\vert$ already reduces much of this. This removal is further strengthened ( without being totally obliterated ) by the cut on $\not \!\!\! E_T$.} over the entire range of DM masses allowed within the kinematic reach of the simulated experiment. For this, we select three values of DM mass $m_\phi$ namely 75, 225, and 325 GeV and we list the observations for various kinematic variables as follows
%%%%%%%%%%%%%%%%%%%%%%
\begin{itemize}
	\item[$\bullet$] For all the kinematic observables (as displayed in Fig.~\ref{fig:distrUS_1TeV}), the variation of the differential cross section for the signal events differs significantly from the SM contribution. The dependence of the signal cross section on DM mass $m_{\phi}$ manifests itself as a limit on the phase space available for the visible particles produced in the final state along with the $\phi$'s. With the increase in DM mass value, the distributions become more distinct, simplifying the problem of determining appropriate selection cuts. More precisely, for the displayed values of DM mass, our first selection cut is $\not \!\! E_T > 100\, \rm GeV$ since the signal events peak for values greater than $100\,\rm GeV$.
	
	\item[$\bullet$] As expected, smaller $m_{\phi}$ values lead to the production of highly boosted jet pair, and the higher values of differential cross section fall in the region corresponding to the smaller $\Delta\phi_{jj}$. With the production of heavier DM particles, the visible particles become less boosted, resulting in a shift in differential cross section distribution towards the higher values of $\Delta\phi_{jj}$. 
	
	\item[$\bullet$] It is evident from the distribution of the energy of the visible pair $E_{j_1}+E_{j_2}$ that the choice of DM mass renders an upper limit on this observable. Theoretically, this upper limit can be understood as a result of the restriction imposed on the invariant mass of visible particles i.e., $\left\vert \left(p_{j_1}+p_{j_2}\right)^2- m_Z^2\right\vert^{\frac{1}{2}}\le 5 \,\Gamma_Z$, where $\Gamma_Z=2.49\,\rm GeV$. This restriction translates to an upper limit on the variable $E_{j_1}+E_{j_2}$ as a function of $m_\phi$: 
	\begin{equation}
	E_{j_1}+E_{j_2} \leq \frac{s+m_Z^2-4m_{\phi}^2}{2\sqrt{s}}.
	\label{eqn:EZcut}
	\end{equation}
	We use this upper limit as our second selection cut to further reduce the SM background.
	Note that, at a linear collider, it is possible to determine the invariant mass of the missing energy-momentum (as long as the initial state radiation effects are not too large) and one would have $m_{invis}^2 \geq 4\,m_{\phi}^2$. However, we prefer to use~\eqref{eqn:EZcut} instead as the l.h.s is easily constructed out of direct observables. The two formulations are equivalent when the jet-pair arise from an on-shell $Z$, which actually constitute the bulk of our signal.
\end{itemize}   
%%%%%%%%%%%%%%
\par To further analyze the signal and SM background events, we choose $\not \!\! E_T > 100\, \rm GeV$ as \textbf{Selection Cut 1} and, the total visible energy given by~\eqref{eqn:EZcut} as \textbf{Selection Cut 2}, which essentially refers to $\phi\,\phi$ production associated with on-shell $Z\to jj\,/ ll$. For the other two channels where the final state has a lepton pair, the observable analogous to that in Eq.~\eqref{eqn:EZcut} would obviously be $E_{l^+}+E_{l^-}$, and a similar selection cut is adopted. Also, analogous to the self-annihilation cross sections in Eqs.~\eqref{sigma_ann_S} and~\eqref{sigma_ann_P}, the DM production cross sections ($\rm e^+\, e^- \rightarrow Z \, \phi \, \phi$) for scalar interactions differ only by a factor of $\mathcal{O}( m^2_f/s)$ from the pseudo-scalar interaction. Hence, our choice of selection cuts is valid for the PS interactions as well.
%%%%%%%%%%%%%%%%%%%%%%
\par We follow a similar approach for the second case with center of mass energy ($\sqrt{s}$) of 500 GeV. While, the selection cut 2 (Eq.~\eqref{eqn:EZcut}) applies to this case as well, we choose a different cut on the missing energy variable i.e. $\not \!\! E_T > 50\, \rm GeV$, appropriate for $\sqrt{s} = 500\, \rm GeV$.
%%%%%%%%%%%%
\section{$\chi^2$ Analysis and Results}
\label{sec:analysisandresults}
%%%%%%%%%%%%%%%
\par The next step of our analysis involves the study of the sensitivity of the cut-off scale $\Lambda$ with the DM mass. Where, as is usual, we set the couplings of the DM particles with the SM fermionic pair to be unity. 
\subsection{$\chi^2$-Statistic}
Assuming a given integrated luminosity, we compare the number of events in a particular bin as expected in the SM and in the presence of New Physics (NP). This allows us to define a $\chi^2$-statistic as 
\begin{eqnarray}
%%%%%%%%%%%%
\chi^2 = \sum_{i=1}^{n}\left ( \frac{n_{i}^{NP}}{\sqrt{n_{i}^{SM+NP}+\delta_{sys}^2(n_{i}^{SM+NP})^2}} \right )^2,
\end{eqnarray} 
%%%%%%%%%%%%%%
where $n$ is the total number of bins, $n_{i}^{NP}$ represents the number of differential events resulting from New Physics and $n_{i}^{SM+NP}$ is the total number of differential events in the $(i)^{th}$ bin in a distribution. The term with $\delta_{sys}$ in the denominator accounts for the systematic uncertainty in our analysis. In contrast to the systematic error of the order of 0.3\% considered in the existing literature~\cite{Behnke:2013xla}, we choose a more conservative value of the order of 1\%. For a given interaction type (S/P), the $\Lambda$ values that can be probed at the ILC for a given value of $\sqrt{s}$ and luminosity are obtained from the most sensitive observable i.e. $\not \!\! E_T$
%%%%%%%%%%%%%%%
\par
\begin{figure*}[ht]
	\centering
	\subfloat{Process 1 : $ e^- e^+ \rightarrow 2 $ jets + $\not \!\! E_T$} \\      
	\includegraphics[width=5.5cm,height=6.5cm,angle=-90]{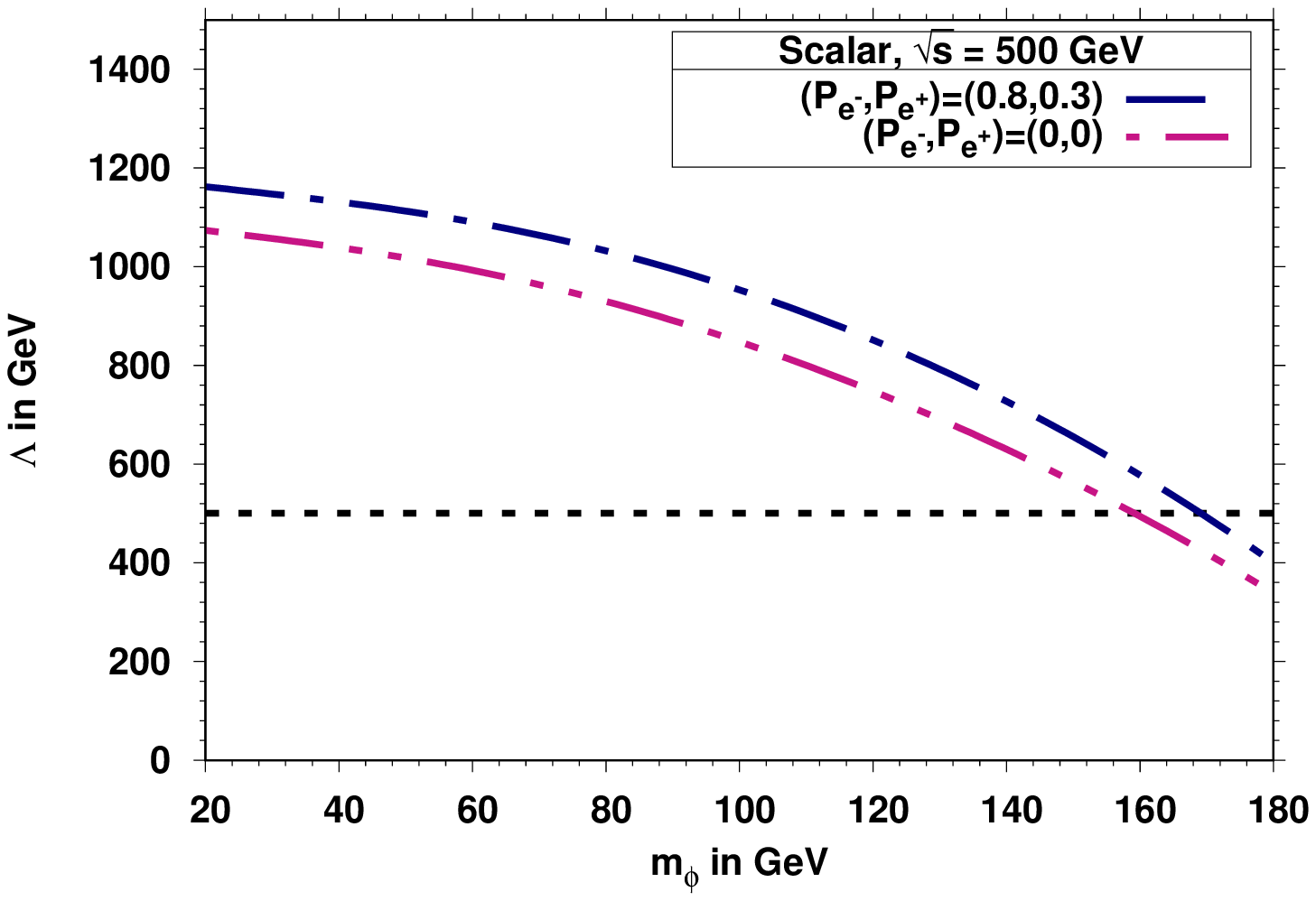}
	\includegraphics[width=5.5cm,height=6.5cm,angle=-90]{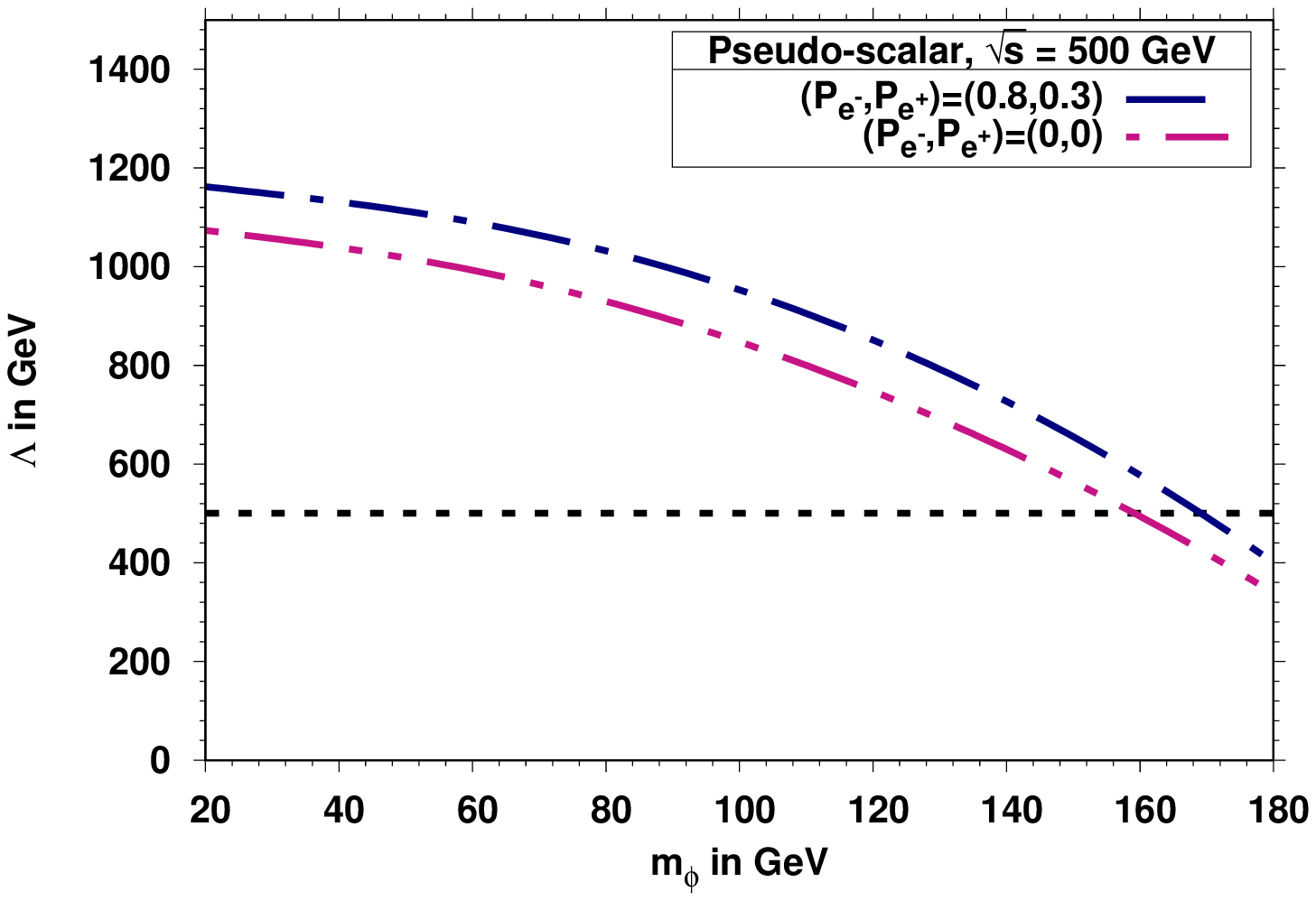}  \\      
	\subfloat{Process 2 : $ e^+ e^- \rightarrow \mu^+ \mu^-$  + $\not \!\! E_T$}\\
	\includegraphics[width=5.5cm,height=6.5cm,angle=-90]{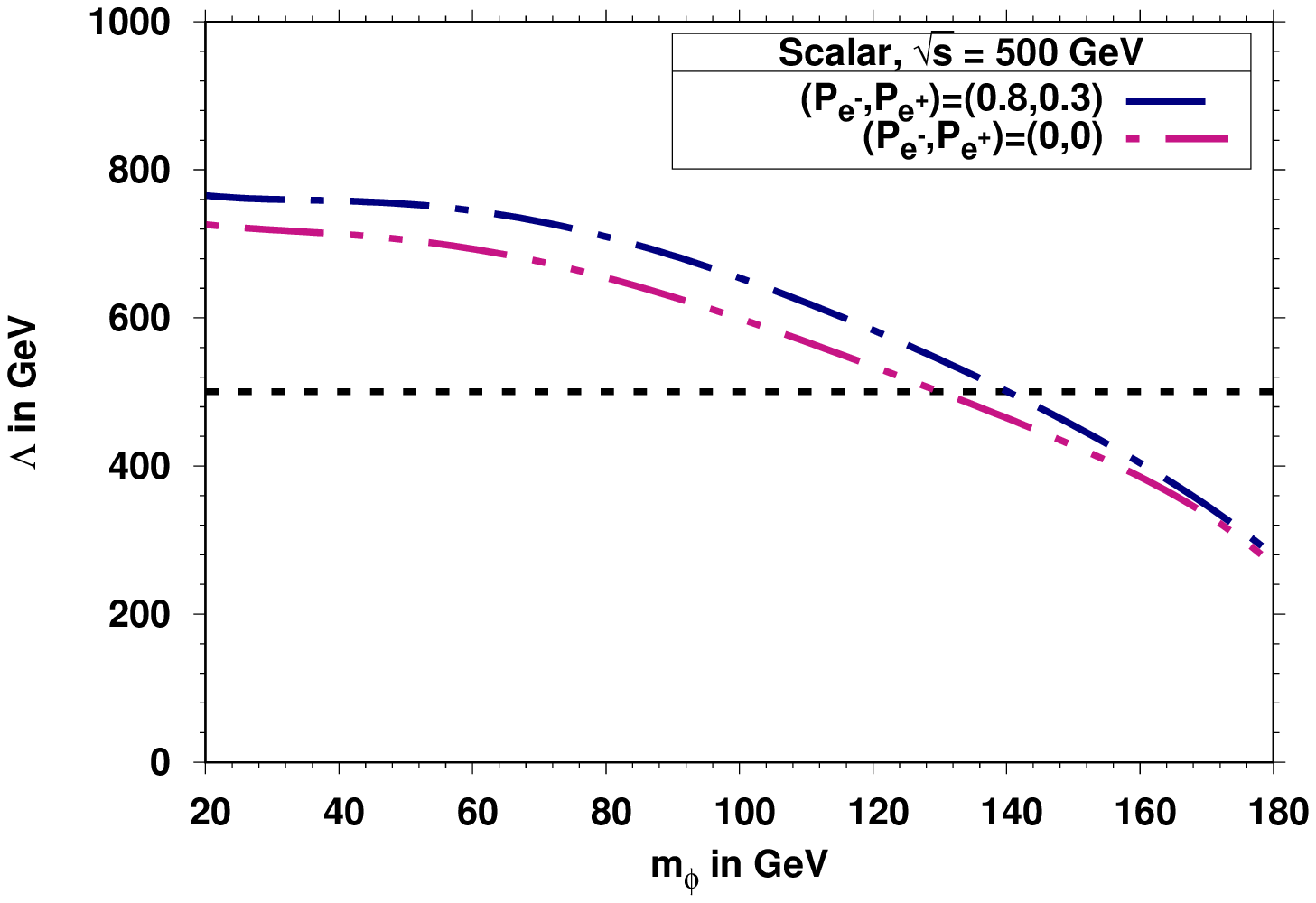}
	\includegraphics[width=5.5cm,height=6.5cm,angle=-90]{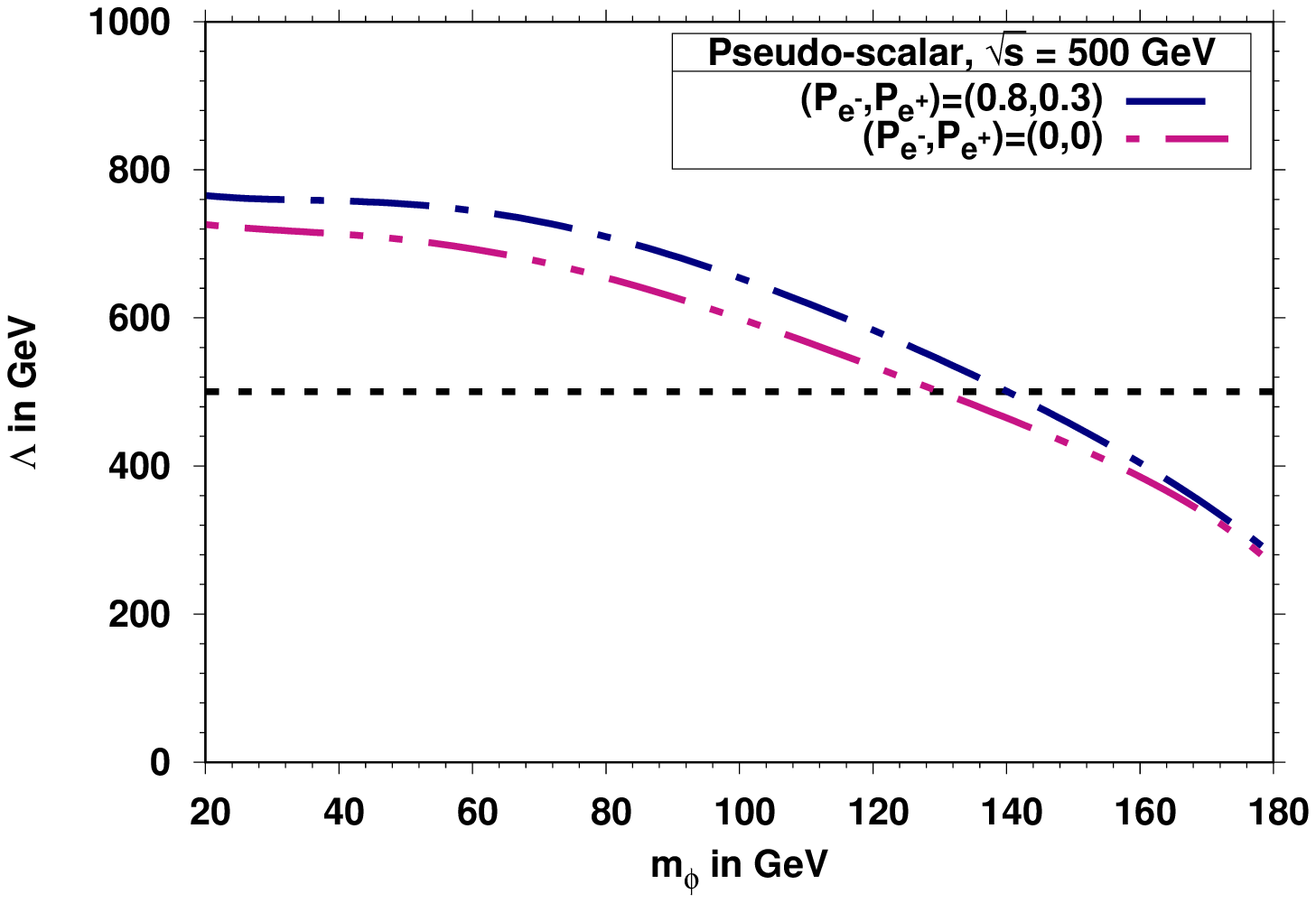}\\
	\subfloat{Process 3 : $ e^+ e^- \rightarrow e^+ e^-$  + $\not \!\! E_T$}\\
	\includegraphics[width=5.5cm,height=6.5cm,angle=-90]{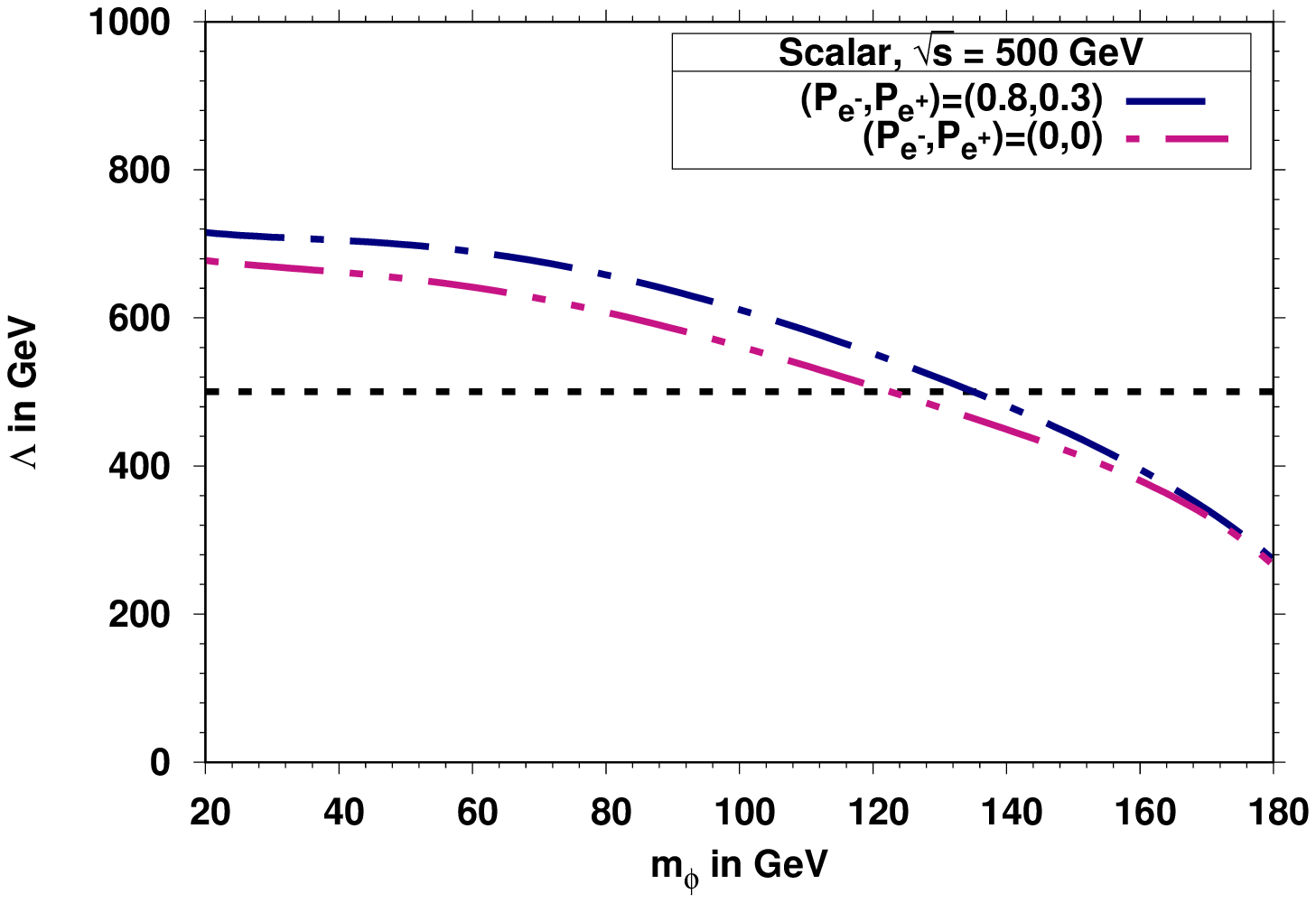}
	\includegraphics[width=5.5cm,height=6.5cm,angle=-90]{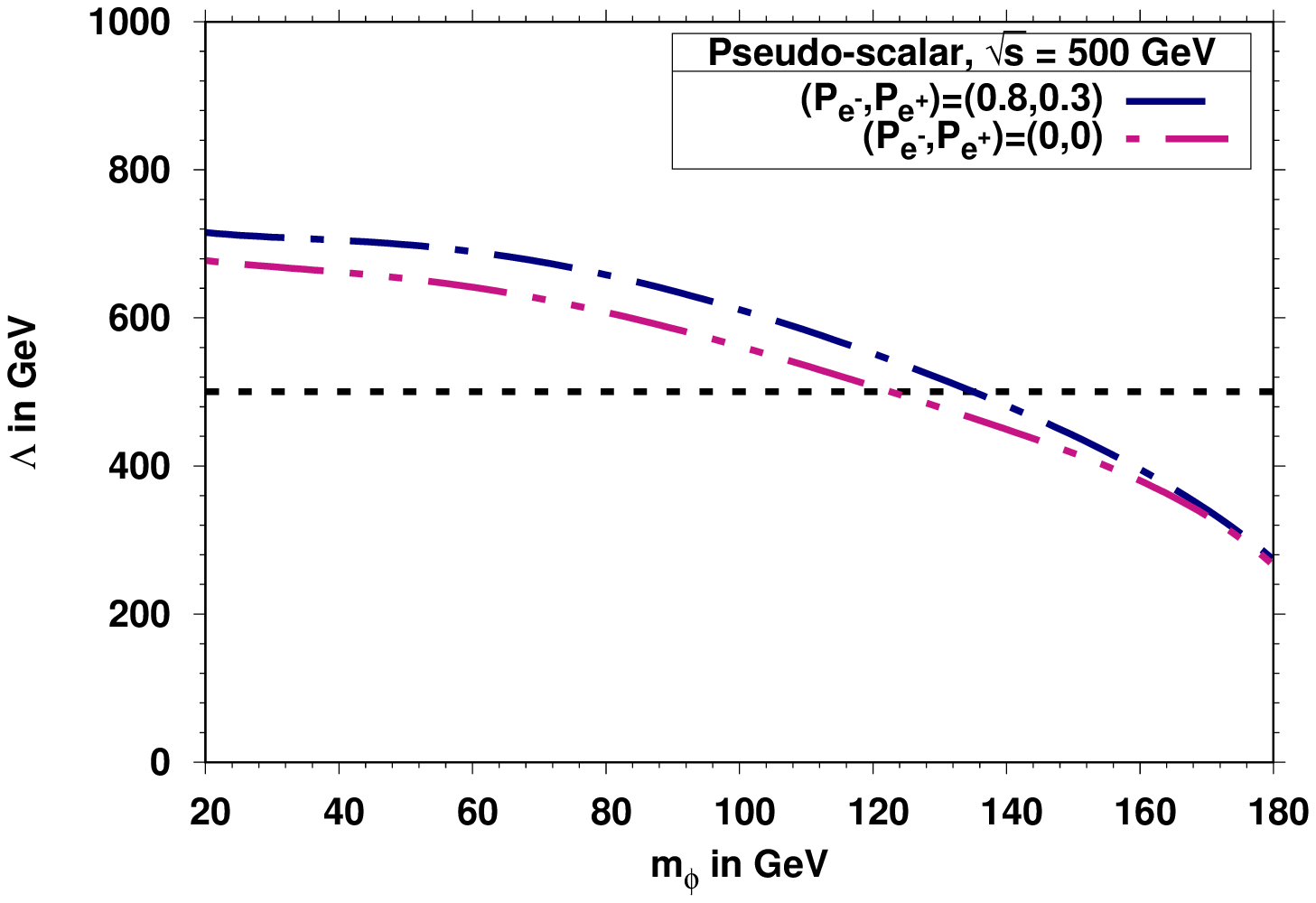}     
	\caption{\small \em{99\% Confidence Level contours in the $m_\phi$ and $\Lambda$ plane  from the  $\chi^2$ analyses of the respective final  visible states at $\sqrt{s}$ = 500 GeV for the collider parameter choice given in Table~\ref{table:accelparam}. The contours at the left  correspond to the scalar interactions of DM with SM leptons ($g_{S}^l$ = 1). The contours at the right correspond to the pseudo-scalar interactions i.e. $g_{P}^l$ = 1. The region below the black dashed line denotes the parameter region for $\Lambda$, where the EFT description breaks down.}}
	\label{fig:contour_500_GeV}
\end{figure*}

\begin{figure*}[ht]
	\centering
	\subfloat{Process 1 : $ e^- e^+ \rightarrow 2 $ jets + $\not \!\! E_T$} \\      
	\includegraphics[width=5.5cm,height=6.5cm,angle=-90]{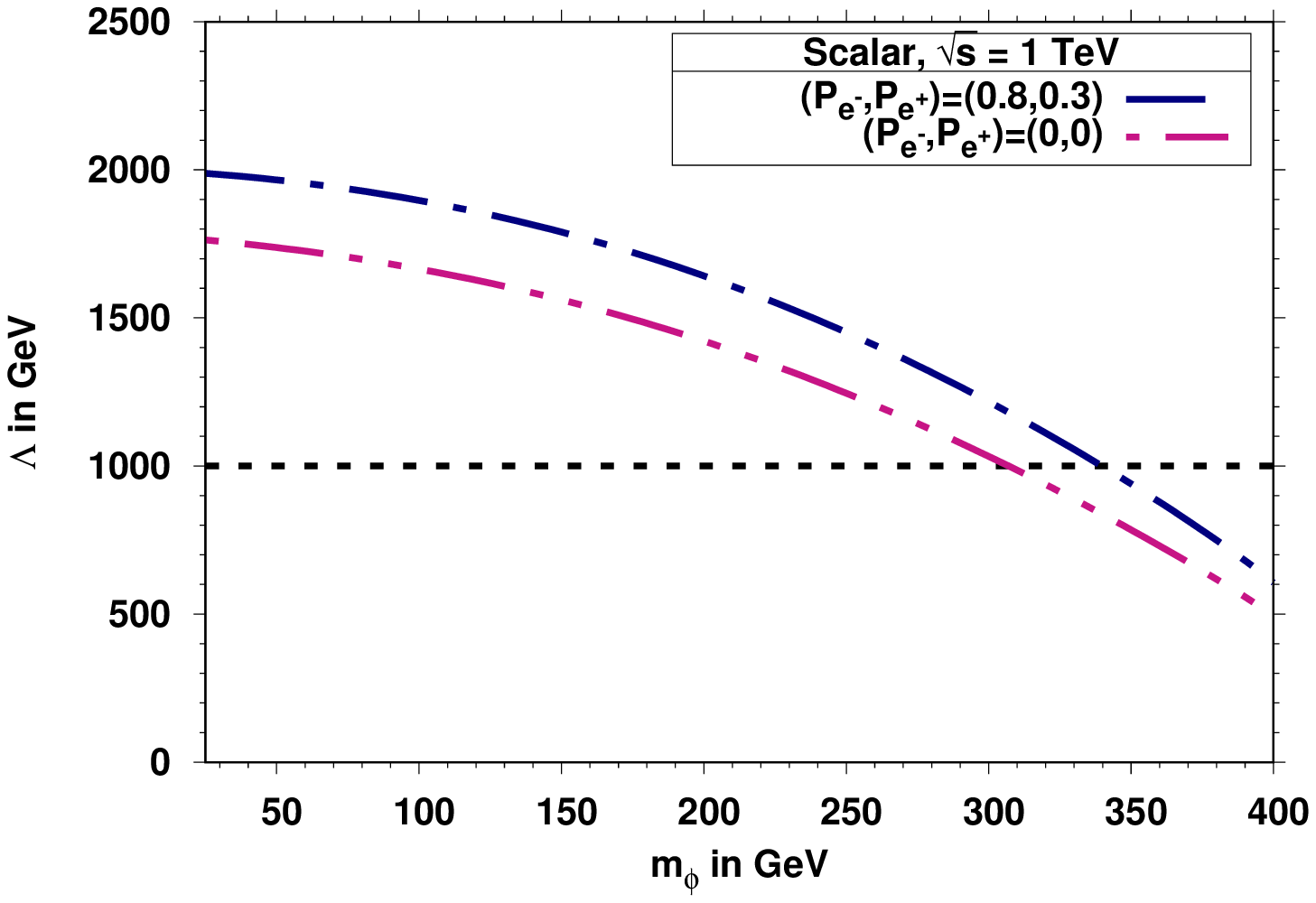}
	\includegraphics[width=5.5cm,height=6.5cm,angle=-90]{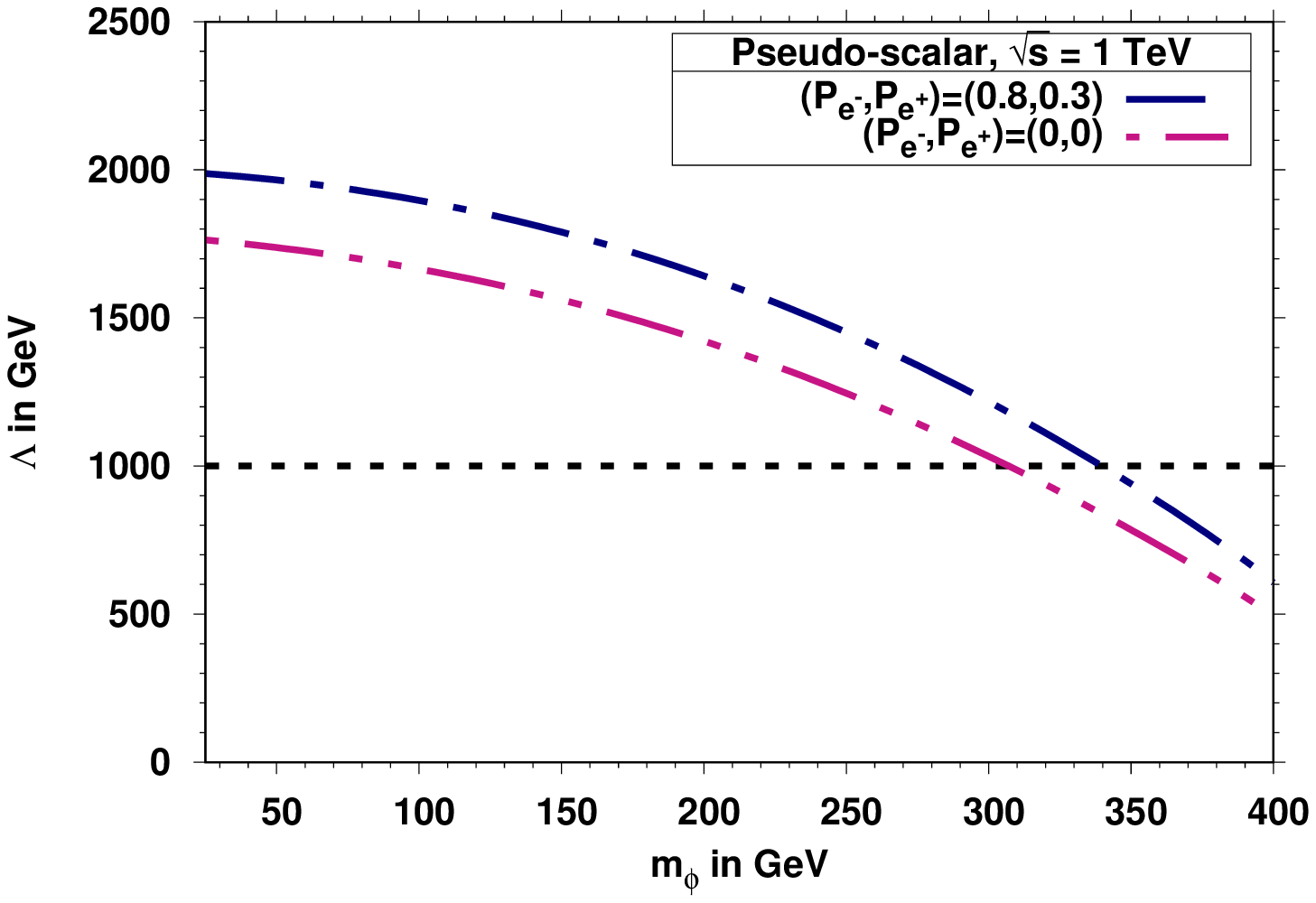}  \\      
	\subfloat{Process 2 : $ e^+ e^- \rightarrow \mu^+ \mu^-$  + $\not \!\! E_T$}\\
	\includegraphics[width=5.5cm,height=6.5cm,angle=-90]{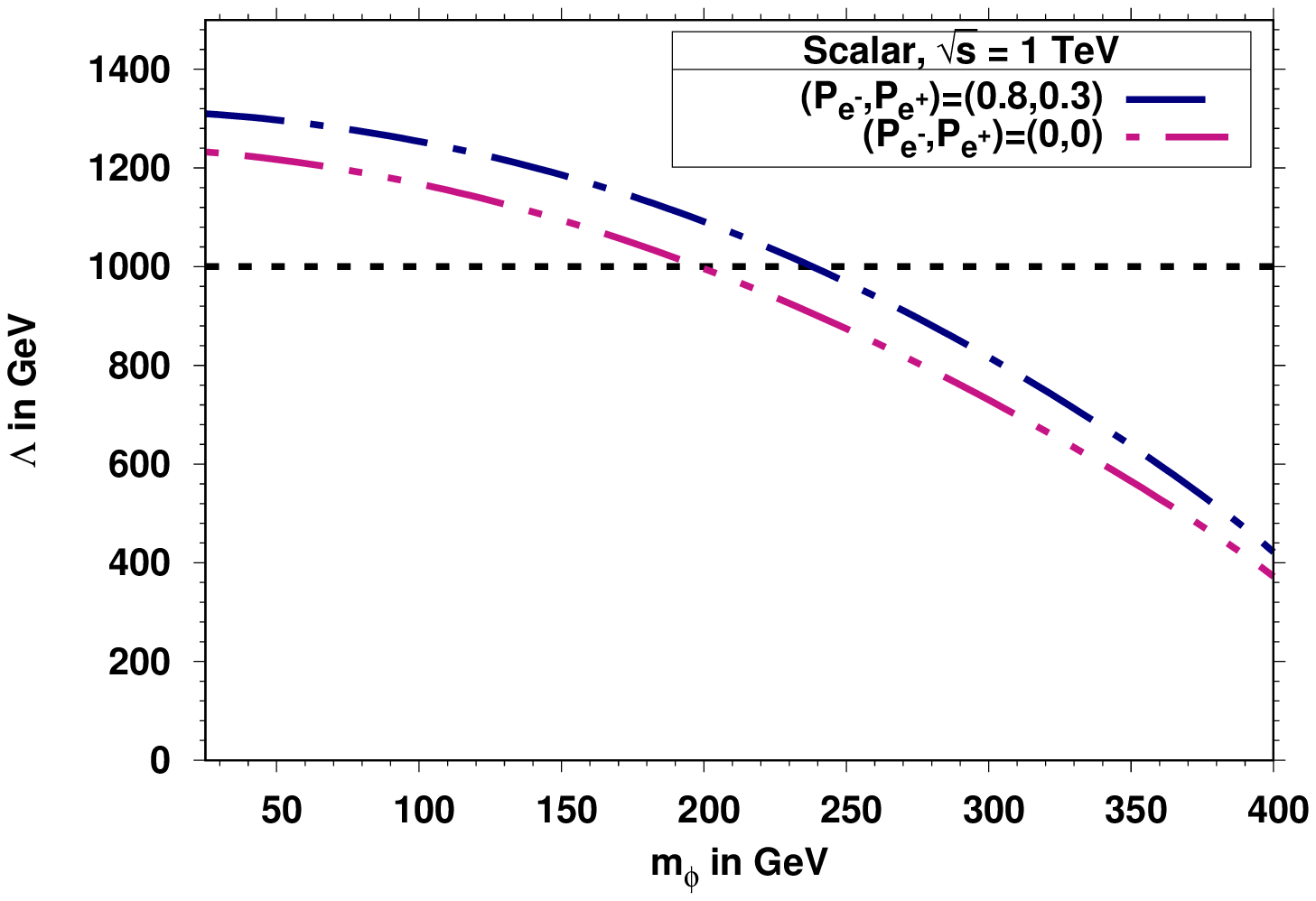}
	\includegraphics[width=5.5cm,height=6.5cm,angle=-90]{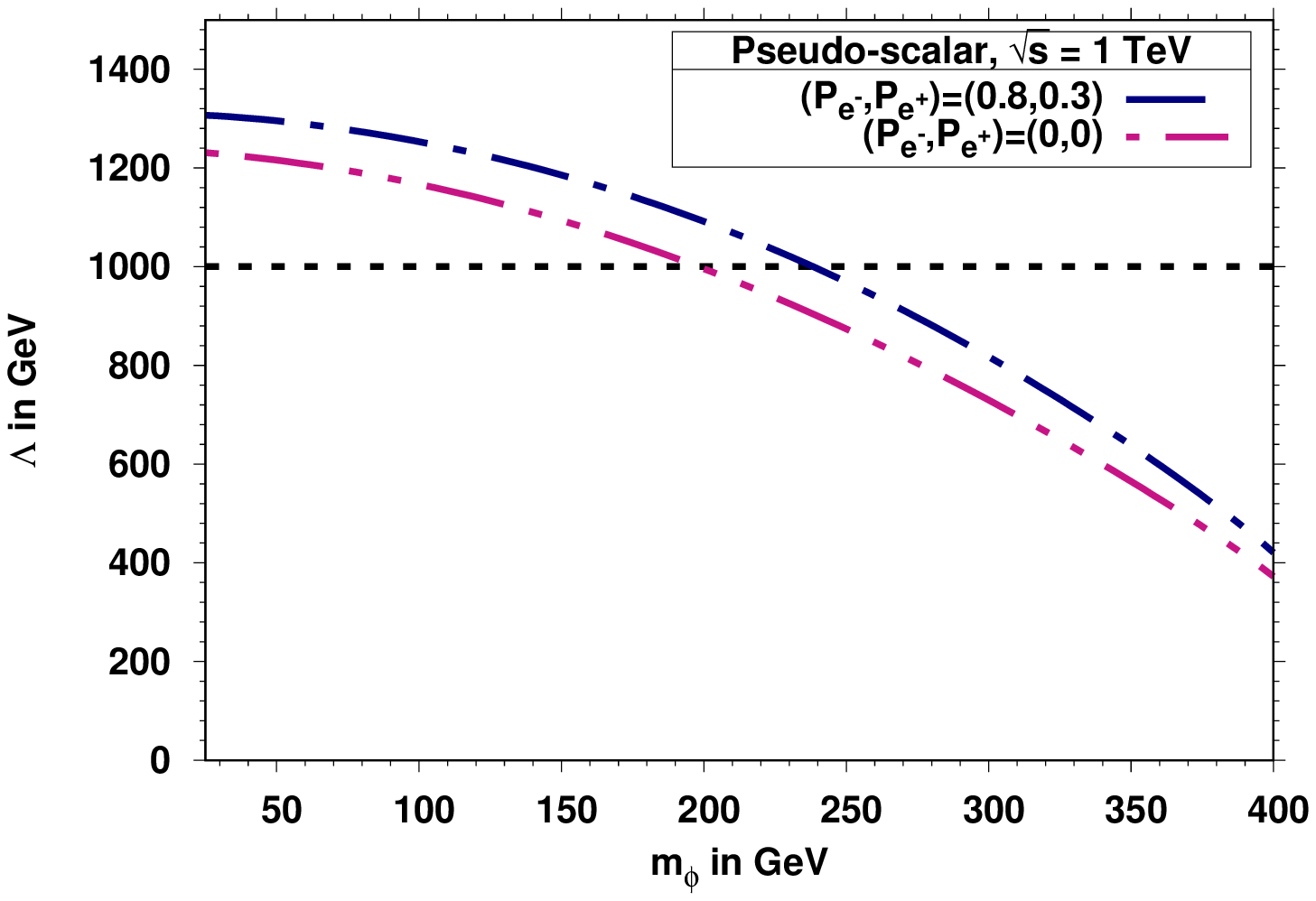}\\
	\subfloat{Process 3 : $ e^+ e^- \rightarrow e^+ e^-$  + $\not \!\! E_T$}\\
	\includegraphics[width=5.5cm,height=6.5cm,angle=-90]{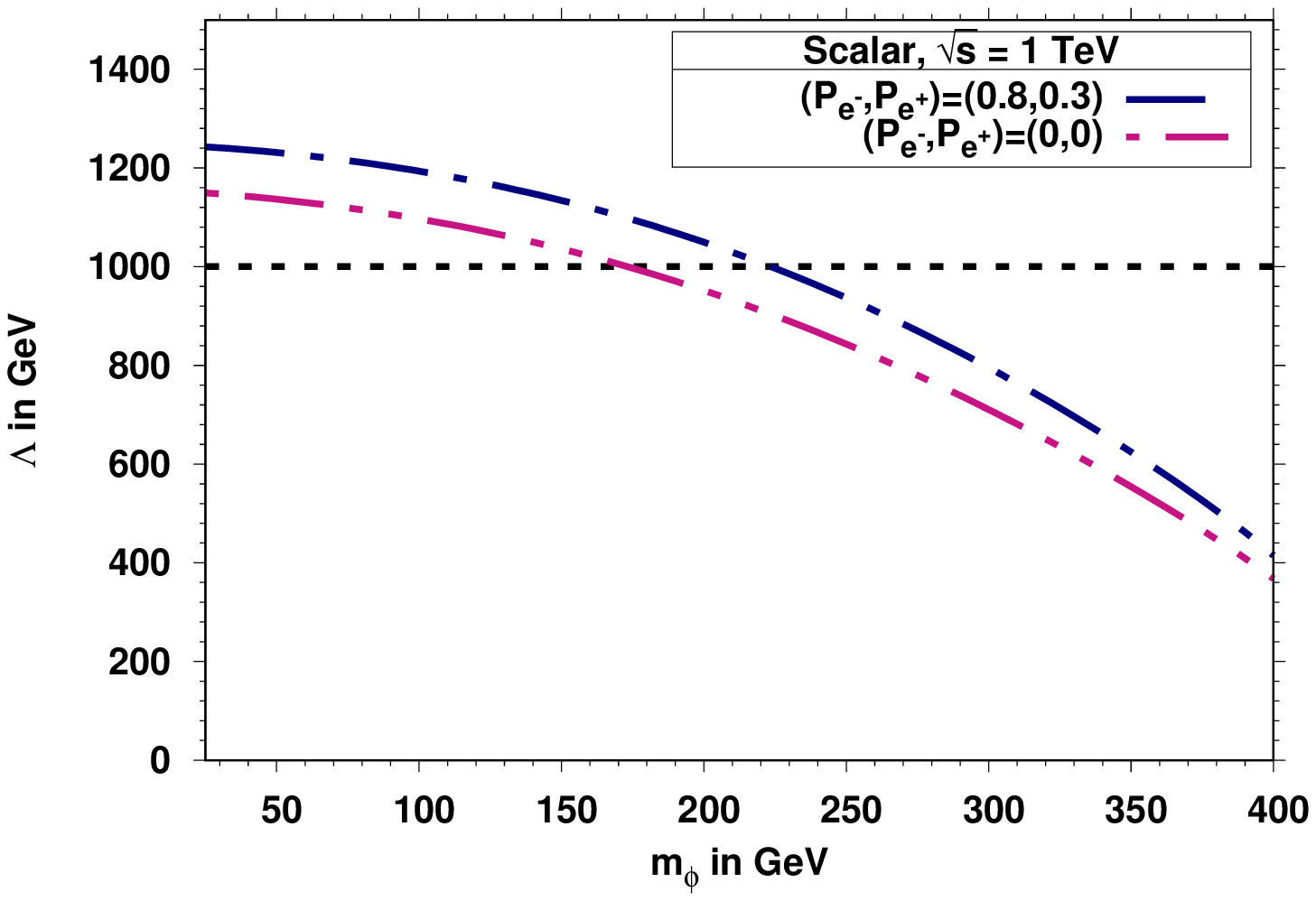}
	\includegraphics[width=5.5cm,height=6.5cm,angle=-90]{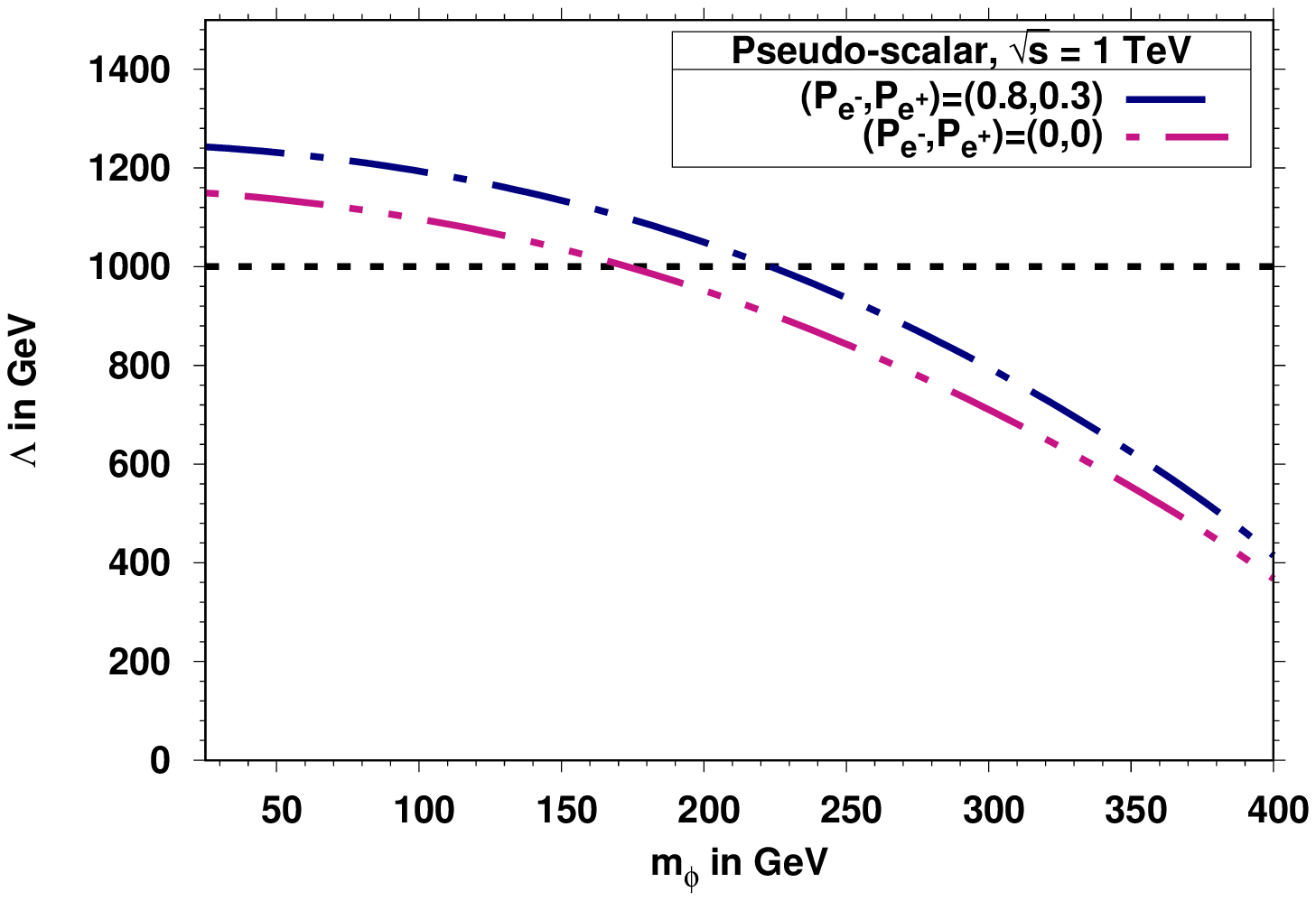}     
	\caption{\small \em{ 99\% Confidence Level contours in the $m_{\phi}$ and $\Lambda$ plane from the $\chi^2$ analyses of the respective final  visible states at $\sqrt{s}$ = 1 TeV for the collider parameter choice given in Table~\ref{table:accelparam}. The contours at the left  correspond to the scalar interactions of DM with SM leptons ($g_{S}^l$ = 1). The contours at the right correspond to the pseudo-scalar interactions i.e. $g_{P}^l$ = 1. The region below the black dashed line denotes the parameter region for $\Lambda$, where the EFT description breaks down.}}
	\label{fig:contour_1_TeV}
\end{figure*}

 In Figs.~\ref{fig:contour_500_GeV} and~\ref{fig:contour_1_TeV}, we present the results of the our $\chi^2$ analysis as 99\% confidence level acceptance contours in the $m_\phi-\Lambda$ plane corresponding to all the three processes. In Table~\ref{table:figureofmerit}, we present the estimates for the 3-$\sigma$ sensitivity reach of the maximum value of the cut-off $\Lambda$ that can be probed in ILC at $\sqrt{s}$ = 500 GeV and 1 TeV with an integrated luminosity $\mathcal{L}$ = 500 fb$^{-1}$ and 1 ab$^{-1}$ respectively. 
%%%%%%%%%%%%
%%%%%%%%%%%%%%%%%%%%%%%%
%%%%%%%%%%%%%%%%
%%%%%%%%%%%%
%%%%%%%%%%%%%%
\begin{table*}[ht]\footnotesize
	\begin{center}
		\begin{tabular}{c||cc|cc||cc|cc}
			\hline\hline
			&\multicolumn{4}{c||}{\bf \underline{Unpolarized}}&\multicolumn{4}{c}{\bf \underline{Polarized}}\\
			
			Interactions    &\multicolumn{2}{c|}{Scalar}&\multicolumn{2}{c||}{Pseudo-scalar}&\multicolumn{2}{c|}{Scalar}&\multicolumn{2}{c}{Pseudo-scalar}\\
			$\sqrt{s}$ in TeV   &0.5 &1&0.5&1 &0.5&1&0.5&1\\
			${\cal L}$ in fb$^{-1}$    &500 &1000&500&1000 &250&500&250&500\\
			$\left(P_{e^-},\, P_{e^+}\right)$    &(0,\,0) &(0,\,0)&(0,\,0)&(0,\,0) &(.8,\,-.3)&(.8,\,-.3)&(.8,\,.3)&(.8,\,.3)\\
			\hline
			$e^+e^-\to \mu^+\mu^-+\not\!\!E_T $ &0.73 &1.23&0.73&1.23&0.76&1.31&0.76&1.31\\
			&&&&&&&&\\ \hline
			$e^+e^-\to e^+e^-+\not\!\!E_T $&0.68&1.15&0.68&1.15&0.72&1.24&0.72&1.24\\
			&&&&&&&&\\ \hline
			$e^+e^-\to j\,j+\not\!\!E_T $ &1.07&1.76&1.07&1.76&1.16&1.99&1.16&1.99\\ 
			&&&&&&&&\\ \hline\hline
		\end{tabular}
	\end{center}
	\caption{\small \em{Estimation of 3$\sigma$ sensitivity reach of the maximum value of the cut-off $\Lambda$ that can be probed in ILC at $\sqrt{s}$ = 500 GeV and 1 TeV with an integrated luminosity $\mathcal{L}$ = 500 fb$^{-1}$ and $\mathcal{L}$ = 1 ab$^{-1}$ respectively, through all visible channels of $e^+e^-\to Z+ \phi\,\phi$.}}
	\label{table:figureofmerit}
\end{table*}
%%%%%%%%%%%%%%%
\clearpage
%%%%%%%%%%%%
\subsection{Effect of Beam Polarization}
%%%%%%%%%%%%%%%
\par The ease of achieving longitudinally polarized electron beams at the ILC motivates us to study the effect of polarized initial electron and positron beams on the sensitivity of $\Lambda$. We, therefore, include the beam polarizations expected at the ILC in our analysis as it can provide an interesting insight in probing the new energy scales. From the helicity conservation requirements, we know that a right-handed particle can only annihilate with a left-handed particle. We can exploit this property to improve our results as t-channel $W$ exchange diagrams contributing to the neutrino background can be suppressed significantly by choosing appropriate polarizations for the electron and positron beams. On the other hand, the rate of pair production of the scalar DM through scalar ($S$) and pseudo-scalar ($P$) interactions will be enhanced for enhances for the case of right-handed electrons and right-handed positrons.
%%%%%%%%%%%
\par The effect of choosing appropriate beam polarization for the inital state leptons, i.e.$+\,80\%$ $e-$ and $+\,30\%$ $e+$, is demonstrated in Figs.~\ref{fig:contour_500_GeV} and~\ref{fig:contour_1_TeV} as the 99\% confidence level contours in the $m_\phi$-$\Lambda$ plane for all possible visible signatures associated with DM pair production at $\sqrt{s} = 500\,\, \rm GeV$ and 1 TeV with an integrated luminosity of 250 and 500 fb$^{-1}$ respectively. For the run scenario corresponding to the center of mass energy of 500 GeV, we observe that the beam polarization increases the $\Lambda$ values marginally and we can probe $\Lambda$ values upto 1.16 TeV at an integrated Luminosity of 250 fb$^{-1}$. However, increasing the center of mass energy ($\sqrt{s}$) to 1 TeV, the beams polarization along with an integrated luminosity of 1 ab$^{-1}$, can enhance the sensitivity of the ILC experiment to probe $\Lambda$ values upto 1.99 TeV.
%%%%%%%%%%%%%%%%%%%%%%%%

\section{Summary}
\label{sec:summary}
%%%%%%%%%%%%%%%%%%%%%%
\par In the search for a solution to the "invisible mass" problem, WIMP candidates with a mass $\sim 10 \, \rm GeV-1\,\rm TeV$, interacting with the visible world with strength nearly of the order of weak interactions, have been studied extensively. Taking motivation from the previous efforts to build leptophilic dark matter models, we assume a scalar DM candidate $\phi$ with significant couplings to leptons only. With this assumption, DM-quark (or, equivalently, DM-gluon) interaction strength would be negligible at the tree level. This immediately renders the direct detection constraints based on the DM-nucleon scattering cross section studies insignificant. Also, the corresponding DM pair-production cross sections (relevant for relic density calculations) at the LHC would be relatively small. ILC is a proposed linear collider, which makes an excellent choice to study the pair production of such a leptophilic DM candidate. In this paper, adopting a simplified model approach, we have performed a detailed analysis to estimate the expected sensitivity of the ILD (one of the two detectors proposed at the ILC) to a scalar leptophilic DM. The recent constraints on the DM contribution to the total energy budget of the universe have been translated and presented as an upper bound on the cut-off scale $\Lambda$ of the effective theory describing the DM-SM fermion interactions in Fig.~\ref{fig:relicdensity}.  

\par We have investigated the DM pair production at the ILC via mono-$Z$ channel, namely $e^+ e^- \rightarrow f\,\bar{f} \phi\,\phi$, where $f$ includes jets, muons, and electrons. With the fully simulated signal and the corresponding SM background for the parameters mentioned in Tabel~\ref{table:accelparam}, we first presented the efficiency of these processes (with the acceptance cuts) in Table~\ref{table:cs}. Next, we have exhibited the effect of appropriate selection cuts based on the study of the normalized one-dimensional differential cross section distributions of various observables on the sensitivity of DM scalar and pseudo-scalar couplings in Fig.~\ref{fig:distrUS_1TeV}. The results obtained from the $\chi^2$ test between the simulated signal and SM background events with respect to the most sensitive kinematic observable are presented in terms of the cut-off scale $\Lambda$. The 99\% confidence level contours in the $m_{\phi}-\Lambda$ plane are displayed in Figs.~\ref{fig:contour_500_GeV} and~\ref{fig:contour_1_TeV}. We find that the suppression of the neutrino background with a specific choice of the initial beam polarization can further enhance the sensitivity of our analyses. We find that with unpolarized initial beams at $\sqrt{s} = 500\, \rm GeV$, we can probe new physics scales $\Lambda$ upto 1.07 TeV with an integrated luminosity of 500 $fb^{-1}$. The improvement with the consideration of some novel effects of beam polarization and an integrated luminosity of 250 $fb^{-1}$ is marginal (1.16 TeV) for this run scenario. However, for the run scenarios focused on a center of mass energy, $\sqrt{s} = 1 \,\rm TeV$, we can probe $\Lambda$ values upto 1.76 TeV with 1 $ab^{-1}$ integrated luminosity. With the appropriate choice of beam polarization, we can enhance the reach of the ILC experiment upto 1.99 TeV for integrated luminosity of 500 $fb^{-1}$.

\par Our study demonstrated that, adopting a simplified model approach, the analysis of the scalar and pseudo-scalar interactions within the mono-$Z$ search channel can provide an insight on the nature of leptophilic DM for the mass range $\sim (20 - 340)$ GeV. It can also provide an idea of the scale associated with this effective theory. Therefore, $Z$-associated DM pair production search channel can prove to be important in probing the cosmologically allowed $m_{\phi}-\Lambda$ regions corresponding to these interactions at the ILC.

%%%%%%%%%%%%%%%%%%%%%%%%%%%%%%%%%%%%%%%%%%%%%%%%%%%%%%%%
\vskip 5mm

\acknowledgments{The author is grateful to Debajyoti Choudhury for useful discussions and reading the manuscript. She would also like to thank Suvam Maharana and Tripurari Srivastava for many useful conversations.}

\end{document}